%% file: paper.tex
\documentclass[twocolumn,showpacs,aps,prl,superscriptaddress]{revtex4}

\usepackage{graphicx} 
\usepackage{dcolumn} 
\usepackage{amsmath} 
\usepackage{epsfig} 
\usepackage{hhline} 
\usepackage{soul} 
\usepackage{tabularx,pifont,multirow} 
\usepackage{colordvi} 
\usepackage{rotate} 
\usepackage[normalem]{ulem}  
 
\input babarsym 

\input symbols

\long\def\inst#1{\par\nobreak\kern 4pt\nobreak 
  {\it #1}\par\vskip 10pt plus 3pt minus 3pt} 
 
\begin{document} 
 
\newcommand{\BABARPubYear}    {12} 
\newcommand{\BABARPubNumber}  {011} 
\newcommand{\SLACPubNumber} {15192} 
\newcommand{\LANLNumber} {1207.5832}

\begin{flushleft} 
\babar-PUB-\BABARPubYear/\BABARPubNumber ~~~~~~~ SLAC-PUB-\SLACPubNumber   \\ 
\end{flushleft}

\title{ 
{ 
\large \bf \boldmath Observation of Time Reversal Violation in the \Bz Meson System 
}
} 

\input authors_apr2012_bad2465.tex

\date{July 20, 2012}
 
\newcommand{\fmv}{\Red } 
\newcommand{\pvp}{\Blue } 
 
\input{abstract} 
\maketitle 
\input{document}

\input acknow_PRL.tex

\input{biblio}
\input{document-extra}

\end{document}

%% file: symbols.tex
\def\CP                {\ensuremath{{C\!P}}\xspace} 
\def\CPT               {\ensuremath{{C\!PT}}\xspace} 
\def\T       {\ensuremath{T}\xspace}

\def\pdf       {\ensuremath{\mathrm{PDF}}\xspace} 
\def\pdfs      {\ensuremath{\mathrm{PDFs}}\xspace}

\newcommand\trule{\rule{0pt}{2.6ex}}

\def\K       {\ensuremath{K}\xspace}

\def\Bminus {\ensuremath{ B_{-} }\xspace} 
\def\Bplus {\ensuremath{ B_{+} }\xspace} 
 
\def\Bflav {\ensuremath{ B_{\rm flav} }\xspace}

\def\dmd {\ensuremath{\Delta m_d}\xspace} 
\def\dt {\ensuremath{ \Delta t }\xspace} 
\def\dttrue {\ensuremath{ \Delta t_{\rm true} }\xspace} 
\def\dtau {\ensuremath{ \Delta \tau }\xspace} 
\def\de {\ensuremath{ \Delta E }\xspace}

\def\SpLpKs  {\ensuremath{ S^+_{\ellp,\KS} }\xspace} 
 
\def\SpLmKs {\ensuremath{ S^+_{\ellm,\KS} }\xspace} 
\def\SmLpKs  {\ensuremath{ S^-_{\ellp,\KS} }\xspace} 
\def\SmLmKs {\ensuremath{ S^-_{\ellm,\KS} }\xspace} 
\def\SpLpKl  {\ensuremath{ S^+_{\ellp,\KL} }\xspace} 
\def\SpLmKl {\ensuremath{ S^+_{\ellm,\KL} }\xspace} 
\def\SmLpKl  {\ensuremath{ S^-_{\ellp,\KL} }\xspace} 
\def\SmLmKl {\ensuremath{ S^-_{\ellm,\KL} }\xspace} 
\def\CpLpKs  {\ensuremath{ C^+_{\ellp,\KS} }\xspace} 
 
\def\CpLmKs {\ensuremath{ C^+_{\ellm,\KS} }\xspace} 
\def\CmLpKs  {\ensuremath{ C^-_{\ellp,\KS} }\xspace} 
\def\CmLmKs {\ensuremath{ C^-_{\ellm,\KS} }\xspace} 
\def\CpLpKl  {\ensuremath{ C^+_{\ellp,\KL} }\xspace} 
\def\CpLmKl {\ensuremath{ C^+_{\ellm,\KL} }\xspace} 
\def\CmLpKl  {\ensuremath{ C^-_{\ellp,\KL} }\xspace} 
\def\CmLmKl {\ensuremath{ C^-_{\ellm,\KL} }\xspace} 
\def\SpBzKs  {\SpLpKs} 
 
\def\SpBzbKs {\SpLmKs} 
\def\SmBzKs  {\SmLpKs} 
\def\SmBzbKs {\SmLmKs} 
\def\SpBzKl  {\SpLpKl} 
\def\SpBzbKl {\SpLmKl} 
\def\SmBzKl  {\SmLpKl} 
\def\SmBzbKl {\SmLmKl} 
\def\CpBzKs  {\CpLpKs} 
 
\def\CpBzbKs {\CpLmKs} 
\def\CmBzKs  {\CmLpKs} 
\def\CmBzbKs {\CmLmKs} 
\def\CpBzKl  {\CpLpKl} 
\def\CpBzbKl {\CpLmKl} 
\def\CmBzKl  {\CmLpKl} 
\def\CmBzbKl {\CmLmKl}

\def\DeltaSpmT {\ensuremath{ \Delta S^\pm_{ \T} }\xspace} 
\def\DeltaCpmT {\ensuremath{ \Delta C^\pm_{ \T} }\xspace}

\def\DeltaSpmCP {\ensuremath{ \Delta S^\pm_{ \CP} }\xspace} 
\def\DeltaCpmCP {\ensuremath{ \Delta C^\pm_{ \CP} }\xspace}

\def\DeltaSpmCPT {\ensuremath{ \Delta S^\pm_{ \CPT} }\xspace} 
\def\DeltaCpmCPT {\ensuremath{ \Delta C^\pm_{ \CPT} }\xspace}

\def\DeltaSmT {\ensuremath{ \Delta S^-_{ \T} }\xspace} 
\def\DeltaCmT {\ensuremath{ \Delta C^-_{ \T} }\xspace}

\def\DeltaSmCP {\ensuremath{ \Delta S^-_{ \CP} }\xspace} 
\def\DeltaCmCP {\ensuremath{ \Delta C^-_{ \CP} }\xspace}

\def\DeltaSmCPT {\ensuremath{ \Delta S^-_{ \CPT} }\xspace} 
\def\DeltaCmCPT {\ensuremath{ \Delta C^-_{ \CPT} }\xspace}

\def\DeltaSpT {\ensuremath{ \Delta S^+_{ \T} }\xspace} 
\def\DeltaCpT {\ensuremath{ \Delta C^+_{ \T} }\xspace}

\def\DeltaSpCP {\ensuremath{ \Delta S^+_{ \CP} }\xspace} 
\def\DeltaCpCP {\ensuremath{ \Delta C^+_{ \CP} }\xspace}

\def\DeltaSpCPT {\ensuremath{ \Delta S^+_{ \CPT} }\xspace} 
\def\DeltaCpCPT {\ensuremath{ \Delta C^+_{ \CPT} }\xspace}

\newcommand{\phm}{\ensuremath{\phantom{-}}} 
 
\def\beq{\begin{equation}} 
\def\eeq{\end{equation}} 
\def\bea{\begin{eqnarray}} 
\def\eea{\end{eqnarray}} 
\def\bq{\begin{quote}} 
\def\eq{\end{quote}} 
\def\ben{\begin{enumerate}} 
\def\een{\end{enumerate}}

\def\twoDLL  {\ensuremath{-2\Delta\ln{\cal L}}\xspace} 
\def\CL {\ensuremath{ \rm CL }\xspace}

%% file: authors_apr2012_bad2465.tex
\author{J.~P.~Lees} 
\author{V.~Poireau} 
\author{V.~Tisserand} 
\affiliation{Laboratoire d'Annecy-le-Vieux de Physique des Particules (LAPP), Universit\'e de Savoie, CNRS/IN2P3,  F-74941 Annecy-Le-Vieux, France} 
\author{J.~Garra~Tico} 
\author{E.~Grauges} 
\affiliation{Universitat de Barcelona, Facultat de Fisica, Departament ECM, E-08028 Barcelona, Spain } 
\author{A.~Palano$^{ab}$ } 
\affiliation{INFN Sezione di Bari$^{a}$; Dipartimento di Fisica, Universit\`a di Bari$^{b}$, I-70126 Bari, Italy } 
\author{G.~Eigen} 
\author{B.~Stugu} 
\affiliation{University of Bergen, Institute of Physics, N-5007 Bergen, Norway } 
\author{D.~N.~Brown} 
\author{L.~T.~Kerth} 
\author{Yu.~G.~Kolomensky} 
\author{G.~Lynch} 
\affiliation{Lawrence Berkeley National Laboratory and University of California, Berkeley, California 94720, USA } 
% ...
\collaboration{The \babar\ Collaboration} 
\noaffiliation

%% file: abstract.tex
\begin{abstract}  
\noindent 
Although \CP violation in the \B meson system has been well established 
by the B factories, there has been no direct observation of time reversal violation. 
The decays of entangled neutral \B mesons into definite flavor states (\Bz or \Bzb), and  
$J/\psi\KL$ or $\ccbar\KS$ final states (referred to as \Bplus or \Bminus), 
allow comparisons between the probabilities of four pairs of \T-conjugated transitions, 
 for example, $\Bzb \to \Bminus$ and $\Bminus \to \Bzb$, 
as a function of the time difference between the two \B decays. 
Using 468 million~\BB pairs produced in $\Upsilon(4S)$ decays collected by the 
\babar~detector at SLAC,  
we measure \T-violating parameters in the time 
evolution of neutral \B mesons, 
yielding 
$\DeltaSpT=-1.37 \pm 0.14~({\rm stat.}) \pm 0.06~({\rm syst.})$ 
and $\DeltaSmT=1.17 \pm 0.18~({\rm stat.}) \pm 0.11~({\rm syst.})$. 
These nonzero results represent 
the first direct observation of \T violation  
through the exchange of initial and final states in transitions that can only be  
connected by a \T-symmetry transformation. 

\end{abstract} 

%% file: document.tex
\indent 
 
The observations of \CP-symmetry breaking, first in neutral \K decays~\cite{ref:christenson:1964} and more 
recently in \B mesons~\cite{ref:mixingInducedCP-Bs,ref:directCP-Bs}, are consistent with the standard model (SM) mechanism  
of the three-family Cabibbo-Kobayashi-Maskawa (CKM) quark-mixing matrix being the dominant source of \CP violation~\cite{ref:CKM:1963:1973}. 
Local Lorentz invariant quantum field theories imply \CPT invariance~\cite{ref:CPTtheorem}, in accordance with all experimental  
evidence~\cite{ref:CPTtests,ref:TestsConservationLaws}. 
Hence, it is expected that the \CP-violating weak interaction also violates time reversal invariance.

To date, the only  
evidence related to \T violation has been found in the neutral \K system, where  
a difference between the probabilities of $\Kz \to \Kzb$ and $\Kzb \to \Kz$ transitions for a given elapsed time  
has been measured~\cite{ref:Angelopoulos}.  
This flavor mixing asymmetry is both \CP- and \T-violating  
(the two transformations lead to the same observation),  
independent of time, 
and requires a nonzero decay width difference $\Delta\Gamma_\K$ between the neutral \K mass  
eigenstates to be observed~\cite{ref:Kabir,ref:Wolfenstein,ref:Wolfenstein2}. 
The dependence with $\Delta\Gamma_\K$ has aroused controversy  
in the interpretation of this observable~\cite{ref:Wolfenstein,ref:Wolfenstein2,ref:Gerber,ref:TestsConservationLaws}. 
In the neutral \B and $\B_s$ systems, where $\Delta\Gamma_d$ and $\Delta\Gamma_s$ are negligible and significantly smaller, respectively, 
the flavor  
mixing 
asymmetry is much more difficult to detect~\cite{ref:TviolationBs}. 
Experiments that could provide direct evidence supporting \T non-invariance,  
without using an observation which also violates \CP, 
involve either nonvanishing expectation values of  
\T-odd observables, or the exchange of initial and final states, which are not \CP conjugates to each other, in the time evolution for transition processes. 
Among the former, there exist upper limits  
for electric dipole moments of the neutron and the electron~\cite{ref:edm}.  
The latter, requiring neutrinos or unstable particles, are particularly difficult to implement. 
 
%
%
%
%
%

In this letter, we report the direct observation of \T violation in the \B meson system, 
through the exchange of initial and final states in transitions that can only be  
connected by a \T-symmetry transformation. 
The method is described in Ref.~\cite{ref:method2012}, based on the concepts proposed in Ref.~\cite{ref:bernabeuPLB-NPB} and further  
discussed in Refs.~\cite{ref:Wolfenstein2,ref:QuinnDiscrete,ref:BernabeuDiscrete}. 
We use a 
data sample of 426~\invfb of integrated luminosity at the \FourS resonance, 
corresponding to $468\times 10^6$~\BB pairs, 
and 45~\invfb at a center-of-mass~(c.m.)~energy 40~\mev below the \FourS,  
recorded by the \babar\ detector~\cite{ref:Aubert:2001tu} at the \pep2 asymmetric-energy $e^+e^-$ collider at  
SLAC. 
The experimental analysis 
exploits  
identical  
reconstruction algorithms, selection criteria,  
calibration techniques, 
and \B meson samples 
to 
our most recent time-dependent \CP asymmetry measurement  
in $\B\to\ccbar K^{(*)0}$ decays~\cite{ref:Aubert:2009yr}, 
with the exception of $\eta_c\KS$ and $\jpsi\Kstarz(\to\KS\piz)$ final states.  
The ``flavor tagging''  
is combined here, for the first time, with the ``\CP tagging''~\cite{ref:bernabeuPLB-NPB}, 
as required for the construction of \T-transformed processes. 
Whereas the  
descriptions 
of the sample composition and time-dependent backgrounds are 
the same as described in Ref.~\cite{ref:Aubert:2009yr}, 
the signal giving access to the \T-violating parameters needs a different data treatment. 
This echoes the fundamental differences between  
observables for \T and \CP symmetry breaking. 
The procedure to determine the \T-violating parameters and their significance is thus novel~\cite{ref:method2012}. 

In the decay of the \FourS, the two \B mesons are in an entangled, 
antisymmetric state, as required by angular momentum conservation for a P-wave particle system. 
This two-body state is usually written in terms of flavor eigenstates,  
such as 
\Bz and \Bzb, 
but can be expressed in terms of any  
linear combinations of \Bz and \Bzb, 
such as the \Bplus and \Bminus states introduced in Ref.~\cite{ref:method2012}. 
They are defined as the neutral \B states  
filtered by the decay to \CP-eigenstates 
$\jpsi\KL$ (\CP-even) and $\jpsi\KS$, with $\KS\to\pi\pi$ (\CP-odd), respectively. 
The \Bplus and \Bminus states are orthogonal to each other 
when there is only one weak phase  
involved in 
the \B decay amplitude, 
as it occurs in \B decays to $\jpsi\Kz$ final states~\cite{ref:CPVreview}, 
and \CP violation in neutral kaons is neglected. 

We select events in which one \B candidate is reconstructed 
in a \Bplus or \Bminus state, and the  
flavor 
of the other \B is identified, 
referred to as flavor identification (ID). 
We generically denote reconstructed final states that  
identify the flavor of the \B as $\ellm X$ for \Bzb and $\ellp X$ for \Bz. 
The notation $(f_1,f_2)$ is used to indicate the flavor or \CP final states  
that are reconstructed at corresponding times $t_1$ and $t_2$, where $t_2>t_1$, 
i.e., $\B_1\to f_1$ is the first decay in the event and $\B_2\to f_2$ is the second decay.  
For later use in Eq.~(\ref{eq:intensitymod}), we define $\dtau=t_2-t_1>0$. 
Once the $\B_1$ state is filtered at time $t_1$, the living partner $\B_2$ is prepared (``tagged'') by entanglement as its orthogonal state. 
The notation  
$\B_2(t_1)\to\B_2(t_2)$  
describes the transition of the \B which  
decays at $t_2$,  
having  
tagged 
its state at $t_1$.  
For example, an event reconstructed in the time-ordered final states $(\ellp X,\jpsi\KS)$ identifies the  
transition $\Bzb\to\Bminus$ for the second \B to decay.  
We  
compare the rate for this transition to  
its \T-reversed $\Bminus\to\Bzb$ 
(exchange of initial and final states) 
by reconstructing the  
final states $(\jpsi\KL,\ellm X)$. 
Any difference in these two rates is  
evidence for \T-symmetry violation. 
There are three other independent comparisons that can be made between  
$\Bplus \to \Bz$ $(\jpsi\KS, \ellp X)$, $\Bzb \to \Bplus$ $(\ellp X,\jpsi\KL)$, and $\Bminus \to \Bz$ $(\jpsi\KL,\ellp X)$ transitions 
and their \T-conjugates, $\Bz \to \Bplus$ $(\ellm X,\jpsi\KL)$, $\Bplus \to \Bzb$ $(\jpsi\KS,\ellm X)$, and $\Bz \to \Bminus$ $(\ellm X,\jpsi\KS)$, 
respectively. 
Similarly, four different \CP(\CPT) comparisons can be made, e.g., between the $\Bzb\to\Bminus$ transition and its \CP(\CPT)-transformed  
$\Bz\to\Bminus$ ($\Bminus\to\Bz$)~\cite{ref:method2012}.

Assuming $\Delta\Gamma_d = 0$, 
each of the eight transitions 
has a general,  
time-dependent decay rate $g_{\alpha,\beta}^\pm(\dtau)$  
given by 
\begin{eqnarray} 
\label{eq:intensitymod} 
e^{-\Gamma_d \dtau} \big\{ 1+ S_{\alpha,\beta}^\pm \sin(\dmd\dtau) + C_{\alpha,\beta}^\pm \cos(\dmd\dtau) \big\}, 
\end{eqnarray} 
where indices $\alpha=\ellp,\ellm$ and $\beta = \KS,\KL$ stand for $\ellp X,\ellm X$ and $\ccbar\KS,\jpsi\KL$ final states, respectively, 
and the  
symbol 
$+$ or $-$ indicates whether the decay to the  
flavor final state $\alpha$  
occurs 
before  
or after  
the decay to the \CP final state $\beta$. 
Here, $\Gamma_d$ is the average decay width, $\dmd$ is the mass difference between the neutral \B mass eigenstates,  
and $C_{\alpha,\beta}^\pm$ and $S_{\alpha,\beta}^\pm$ are  
model independent 
coefficients. 
The sine term, 
expected to be large in the SM, 
results from the interference between direct decay of the neutral \B to the $\jpsi\Kz$ final state and decay after  
\mbox{\Bz-\Bzb} 
oscillation,  
while the cosine term arises from the interference between decay amplitudes with different weak and strong phases, 
and is expected to be negligible~\cite{ref:CPVreview}. 
\T violation would manifest itself through differences between the $S_{\alpha,\beta}^\pm$ or $C_{\alpha,\beta}^\pm$ values for \T-conjugated processes, 
for example between \SpLpKs and \SmLmKl.  

In addition to $\jpsi\KS$, \Bminus states are reconstructed through the $\psi(2S)\KS$ and $\chi_{c1}\KS$ final states 
(denoted generically as $\ccbar\KS$),  
with $\jpsi,\psi(2S) \to e^+e^-,\mu^+\mu^-$, $\psi(2S)\to\jpsi\pip\pim$, $\chi_{c1}\to\jpsi\gamma$, and $\KS\to\pip\pim,\piz\piz$ 
(the latter only for $\jpsi\KS$). 
\Bplus states are identified through $\jpsi\KL$. 
The $\jpsi\KL$ candidates are characterized by the difference  
\de between the reconstructed energy of the \B and the beam energy in the $e^+e^-$ c.m. frame, $E_{\rm beam}^*$, 
while for the $\ccbar\KS$ modes we use the beam-energy substituted invariant mass $\mes = \sqrt{(E_{\rm beam}^*)^2-(p_\B^*)^2}$, 
where $p_\B^*$ is the \B momentum in the c.m. frame. 

The flavor ID of the other neutral \B meson in the event, not associated with the reconstructed \Bplus or \Bminus, 
is made 
on the basis of the charges of  
prompt leptons, 
kaons, pions from $D^*$ mesons, 
and high-momentum charged particles.  
These flavor ID inputs are 
combined using a neural network (NN), 
trained with Monte Carlo (MC) simulated data. 
The output of the NN is then divided into six hierarchical, mutually exclusive  
flavor categories of  
increasing misidentification (misID)  
probability $w$. 
Events for which the  
NN output indicates 
very low discriminating power are excluded from further analysis. 
We determine the  
signed  
difference of proper time  
$\dt=t_{\beta}-t_{\alpha}$ 
between the two \B decays from the measured separation  
of the decay vertices along the collision axis.  
Events are accepted if the reconstructed $|\dt|$ and its estimated uncertainty, $\sigma_{\dt}$, are 
lower than $20$~\ps and $2.5$~\ps, respectively. 
The performances of the flavor ID and \dt reconstruction algorithms are evaluated 
by using a large sample of flavor-specific neutral \B decays to $D^{(*)-}[\pip,\rho(770)^+,a_1(1260)^+]$ and $\jpsi \Kstarz(\to \Kp\pim)$ final  
states (referred to as \Bflav  
sample). 
The \dt resolution function is the same as in Ref.~\cite{ref:Aubert:2009yr} except that all  
Gaussian offsets and widths are modeled to be proportional to $\sigma_{\dt}$.

\begin{figure}[htb] 
\begin{center} 
\begin{tabular}{cc} 
\includegraphics[width=0.23\textwidth]{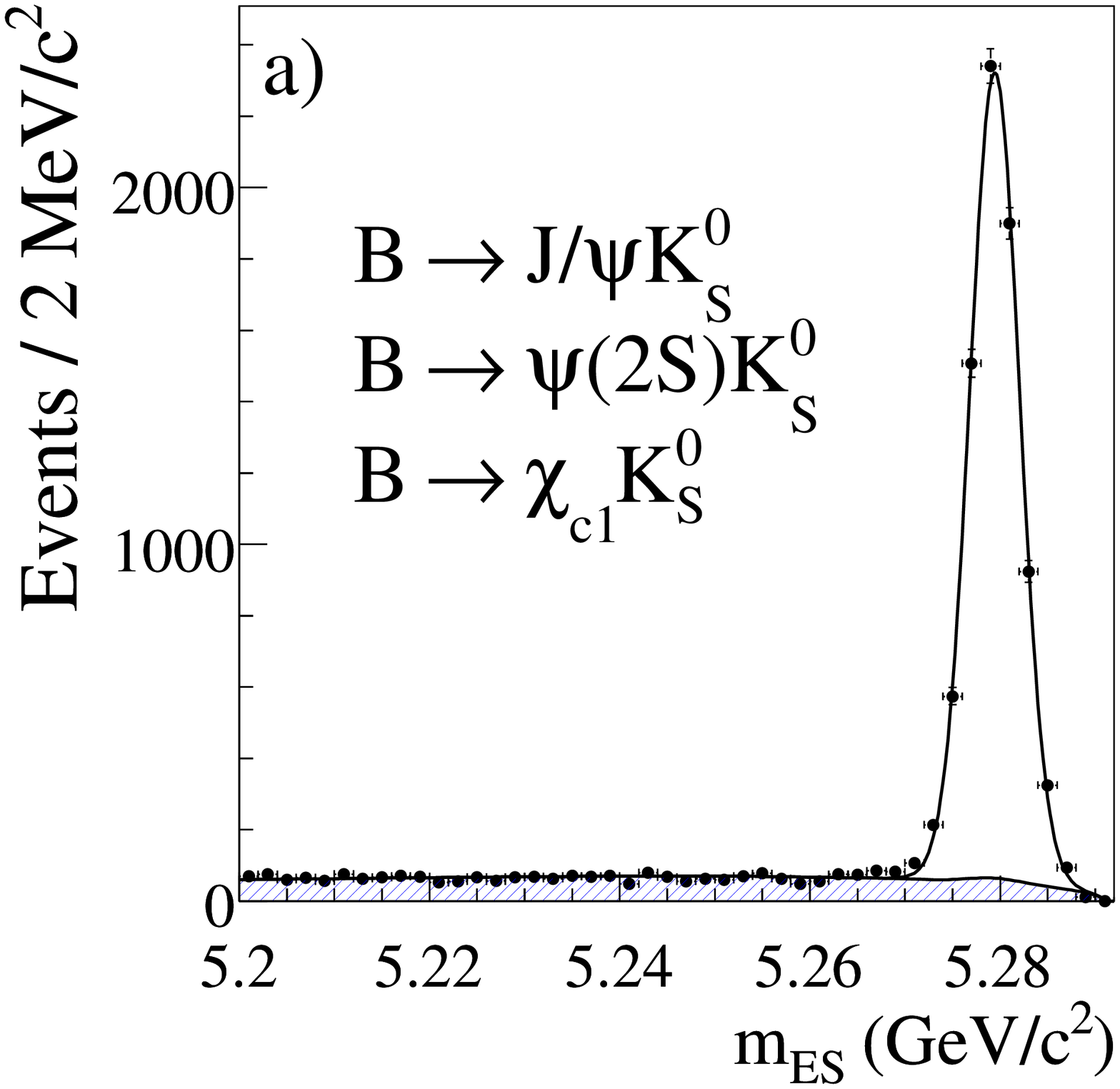} 
\includegraphics[width=0.23\textwidth]{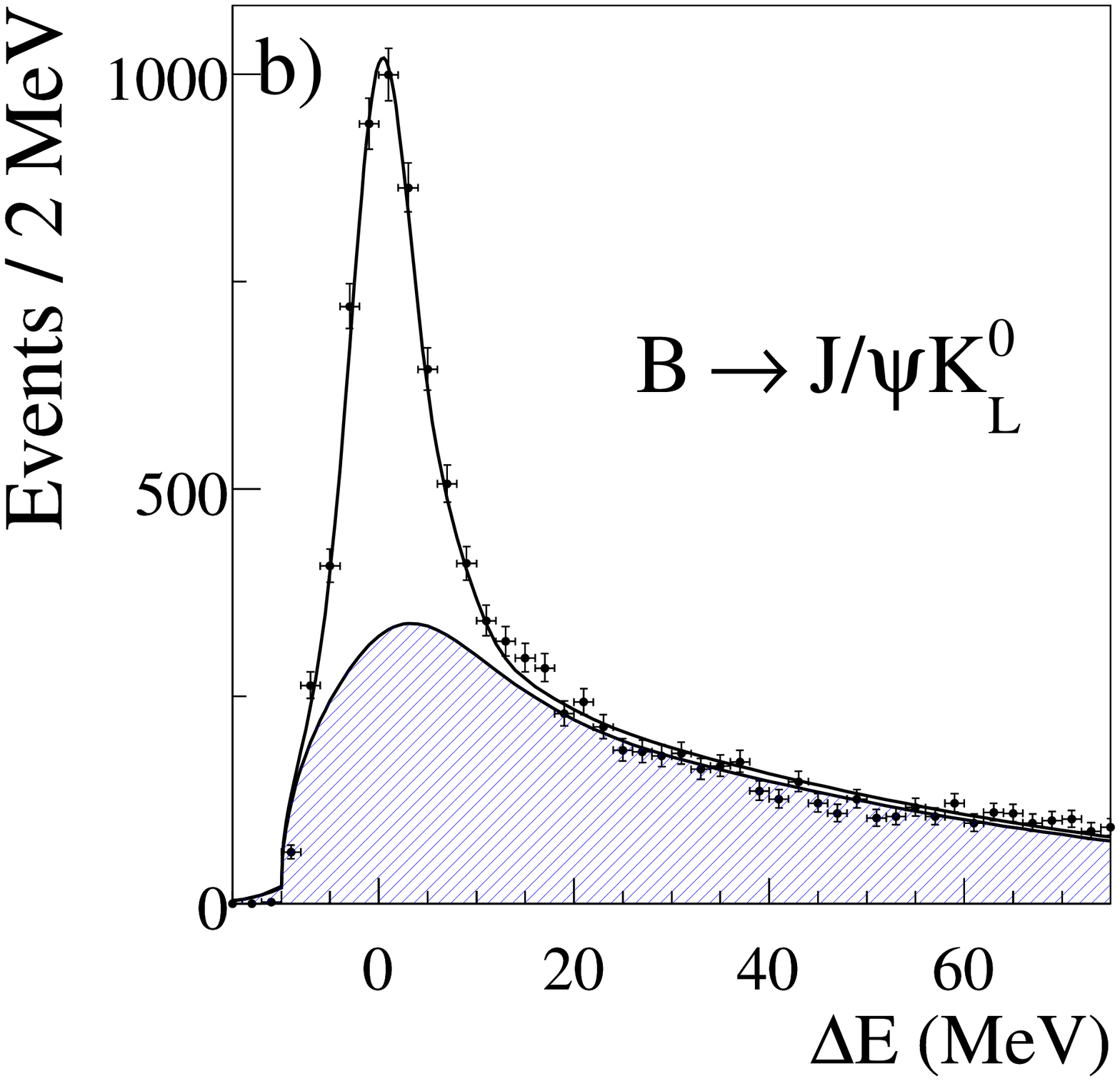} 
\end{tabular} 
\caption{\label{fig:datasample} (color online). 
Distributions of (a) \mes and (b) \de for the neutral \B decays reconstructed in the $\ccbar\KS$ and $\jpsi\KL$ final states, 
respectively, after flavor ID and vertexing requirements. In each plot, the shaded region is the estimated background contribution. 
The two samples of events are identical to those used in our most recent \CP-violation study~\cite{ref:Aubert:2009yr}, 
but excluding $\eta_c\KS$ and $\jpsi\Kstarz(\to\KS\piz)$ final states. 
} 
\end{center} 
\end{figure}

The composition of the final sample is determined  
through fits to the \mes and \de distributions, using parametric forms and distributions extracted from  
MC simulation and  
dilepton mass sidebands in data to describe the signal and background components.   
Figure~\ref{fig:datasample} shows the \mes and \de data distributions for events that satisfy 
the flavor ID and vertexing requirements, overlaid with the fit projections. 
The final sample contains 7796 $\ccbar\KS$ events, with purities in the signal region ($5.27< \mes < 5.29$~\gevcc) 
ranging between 87\% and 96\%, and 5813 $\jpsi\KL$ events, with a purity of 56\% in the $|\de|<10$~\mev region. 
 
We perform a simultaneous, unbinned maximum likelihood fit to the \dt distributions  
for flavor identified $\ccbar\KS$ and $\jpsi\KL$ events, split by flavor category. 
The signal probability density function (\pdf) is~\cite{ref:method2012}  
\begin{eqnarray} 
{\cal H}_{\alpha,\beta}(\dt) & \propto  & g_{\alpha,\beta}^+( \dttrue) H( \dttrue) \otimes \mathcal{R}(\delta t;\sigma_{\dt})+ ~~~~ \label{eq:intensitydt} 
\\ 
                             &          & g_{\alpha,\beta}^-(-\dttrue) H(-\dttrue) \otimes \mathcal{R}(\delta t;\sigma_{\dt}) , \nonumber 
\end{eqnarray} 
where \dttrue is the signed difference of proper time between the two \B decays in the limit of perfect \dt reconstruction, 
$H$ is the Heaviside step function,  
$\mathcal{R}(\delta t;\sigma_{\dt})$ with $\delta t=\dt - \dttrue$ is the resolution function,   
and $g_{\alpha,\beta}^\pm$ are given by Eq.~(\ref{eq:intensitymod}). 
Note that \dttrue is equivalent to $\dtau$ ($-\dtau$) when a true flavor (\CP) tag occurs. 
Because of the convolution with the resolution function, 
the distribution for \mbox{$\dt>0$} contains predominantly true flavor-tagged events,  
with  
contribution from true \CP-tagged events at low \dt, and  
conversely 
for \mbox{$\dt<0$}.  
Mistakes in the flavor ID algorithm 
mix correct and incorrect flavor assignments,  
and dilute the \T-violating asymmetries by a factor of approximately  
$(1-2w)$. 
Backgrounds are accounted for by adding terms to  
Eq.~(\ref{eq:intensitydt})~\cite{ref:Aubert:2009yr}. 
Events are assigned signal and background probabilities based on the \mes or \de distributions, 
for $\ccbar\KS$ or $\jpsi\KL$ events, respectively.

A total of 27 parameters are varied in the likelihood fit: 
eight pairs of $(S_{\alpha,\beta}^\pm, C_{\alpha,\beta}^\pm)$ coefficients for the signal, 
and 11 parameters describing possible \CP and \T violation in the background.  
All remaining signal and background parameters 
are fixed to values taken from the \Bflav sample,  
\jpsi-candidate sidebands in $\jpsi\KL$,  
world averages for $\Gamma_d$ and \dmd~\cite{ref:pdg2010}, or MC simulation~\cite{ref:Aubert:2009yr}. 
From the 16 signal coefficients~\cite{ref:epaps}, we construct six pairs of independent asymmetry parameters  
$(\DeltaSpmT, \DeltaCpmT)$, $(\DeltaSpmCP, \DeltaCpmCP)$, and $(\DeltaSpmCPT, \DeltaCpmCPT)$,  
as shown in Table~\ref{tab:results}. 
The \T-asymmetry parameters have the advantage that \T-symmetry breaking 
would directly manifest itself   
through any nonzero value of \DeltaSpmT or \DeltaCpmT, or any 
difference between \DeltaSpmCP and \DeltaSpmCPT, or between \DeltaCpmCP and \DeltaCpmCPT 
(analogously for \CP- or \CPT-symmetry breaking). 
The measured values for the asymmetry parameters are reported in Table~\ref{tab:results}. 
There is another two times three pairs of \T-, \CP-, and \CPT-asymmetry parameters, 
but they are not independent and can be derived from Table~\ref{tab:results} or Ref.~\cite{ref:epaps}.

\begin{table}[htb!]        
\begin{center} 
\caption{ 
Measured values of the \T-, \CP-, and \CPT-asymmetry parameters, defined as the 
differences in $S_{\alpha,\beta}^\pm$ and $C_{\alpha,\beta}^\pm$ between symmetry-transformed transitions. 
The values of reference coefficients are also given at the bottom. 
The first uncertainty is statistical and the second systematic. 
The indices \ellm, \ellp, \KS, and \KL stand for reconstructed final states that  
identify the \B meson as \Bzb, \Bz, \Bminus, and \Bplus, respectively. 
\label{tab:results}} 
\begin{ruledtabular} 
       \begin{tabular}{ l  c }		 
         Parameter & Result \\ \hline 
\trule   \DeltaSpT  = \SmLmKl $-$ \SpLpKs    &   $ -1.37\pm 0.14\pm 0.06$ \\   
\trule   \DeltaSmT   = \SpLmKl $-$ \SmLpKs   &   $ \phm1.17\pm 0.18\pm 0.11$ \\   
\trule   \DeltaCpT    = \CmLmKl $-$ \CpLpKs  &   $ \phm0.10\pm 0.14\pm 0.08$ \\ 
\trule   \DeltaCmT   = \CpLmKl $-$ \CmLpKs   &   $ \phm0.04\pm 0.14\pm 0.08$ \\ [0.07in] \hline 
\trule   \DeltaSpCP = \SpLmKs $-$ \SpLpKs    &   $ -1.30\pm 0.11\pm 0.07$ \\  
\trule   \DeltaSmCP  = \SmLmKs $-$ \SmLpKs   &   $ \phm1.33\pm 0.12\pm 0.06$ \\  
\trule   \DeltaCpCP   = \CpLmKs $-$ \CpLpKs  &   $ \phm0.07\pm 0.09\pm 0.03$ \\ 
\trule   \DeltaCmCP  = \CmLmKs $-$ \CmLpKs   &   $ \phm0.08\pm 0.10\pm 0.04$ \\ [0.07in] \hline 
\trule   \DeltaSpCPT = \SmLpKl $-$ \SpLpKs   &   $ \phm0.16\pm 0.21\pm 0.09$ \\   
\trule   \DeltaSmCPT = \SpLpKl $-$ \SmLpKs   &   $ -0.03\pm 0.13\pm 0.06$ \\ 
\trule   \DeltaCpCPT = \CmLpKl $-$ \CpLpKs   &   $ \phm0.14\pm 0.15\pm 0.07$ \\ 
\trule   \DeltaCmCPT   = \CpLpKl $-$ \CmLpKs &   $ \phm0.03\pm 0.12\pm 0.08$ \\ [0.07in] \hline 
\trule   \SpLpKs     &   $ \phm0.55\pm 0.09\pm 0.06$ \\ 
\trule   \SmLpKs     &   $ -0.66\pm 0.06\pm 0.04$ \\ 
\trule   \CpLpKs     &   $ \phm0.01\pm 0.07\pm 0.05$ \\  
\trule   \CmLpKs     &   $ -0.05\pm 0.06\pm 0.03$ \\ [0.05in] 
       \end{tabular} 
\end{ruledtabular} 
    \end{center}	 
  \end{table}

\begin{figure}[htb] 
\begin{center} 
\begin{tabular}{cc} 
  \includegraphics[width=0.23\textwidth]{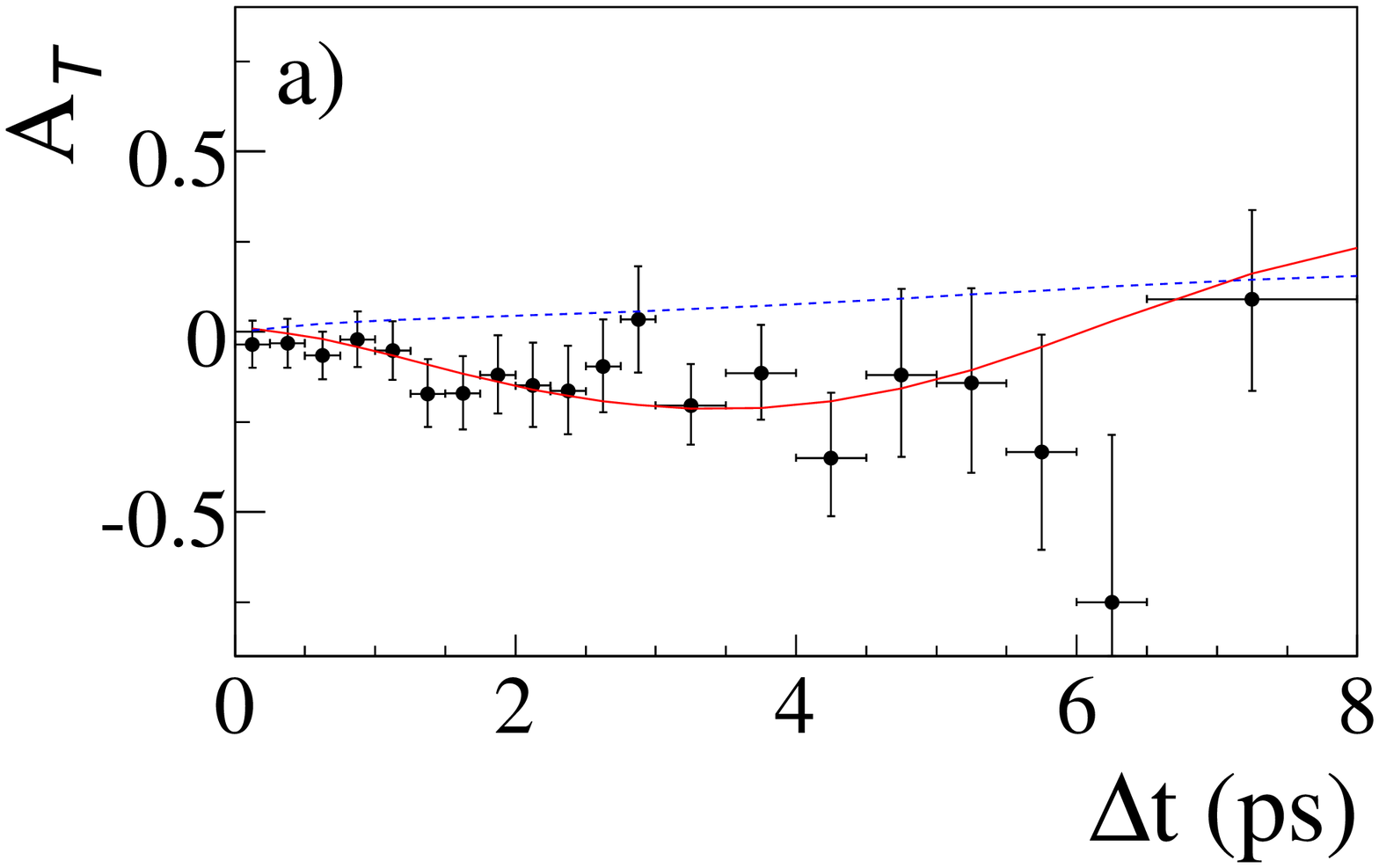} &  
  \includegraphics[width=0.23\textwidth]{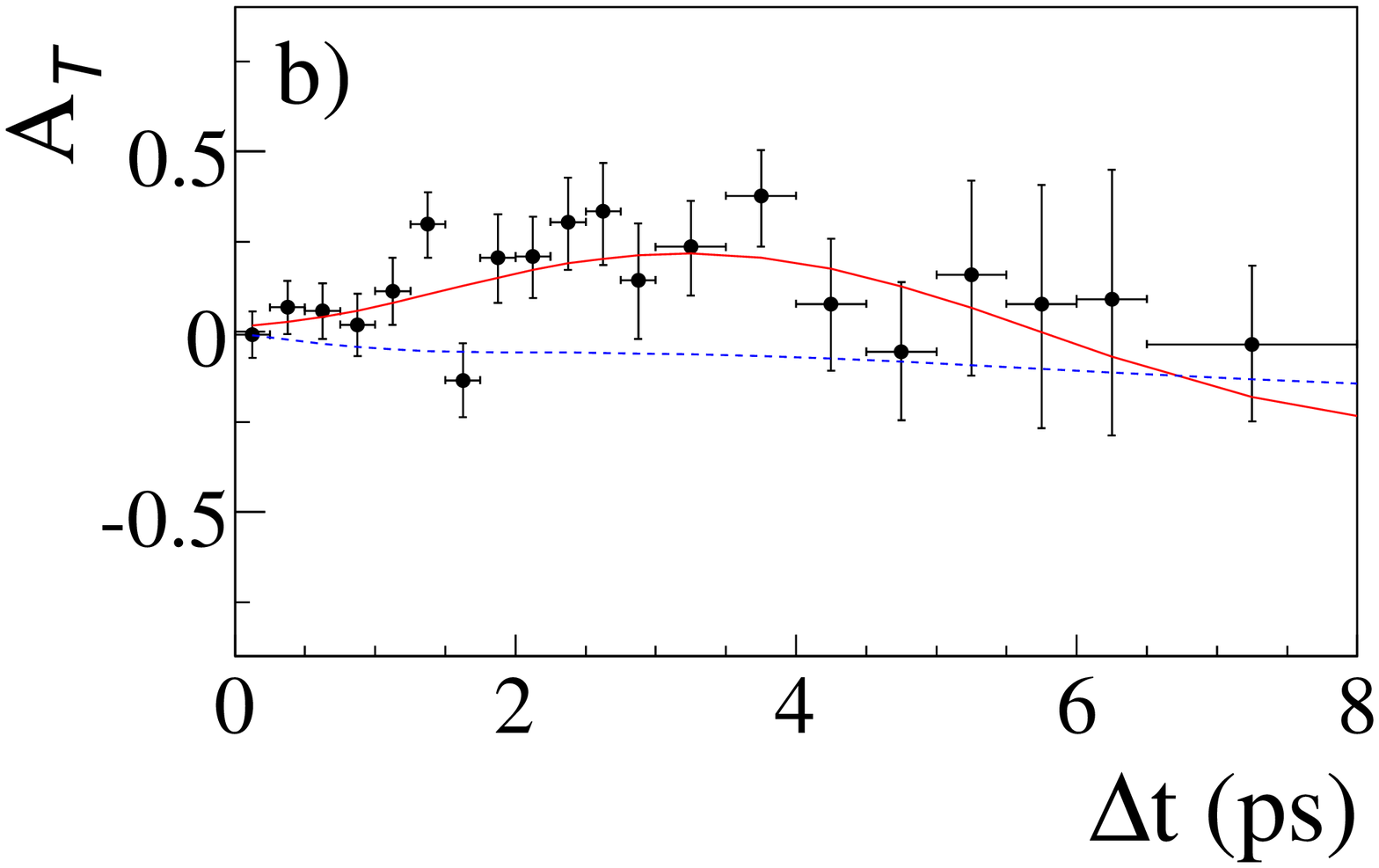} \\ 
  \includegraphics[width=0.23\textwidth]{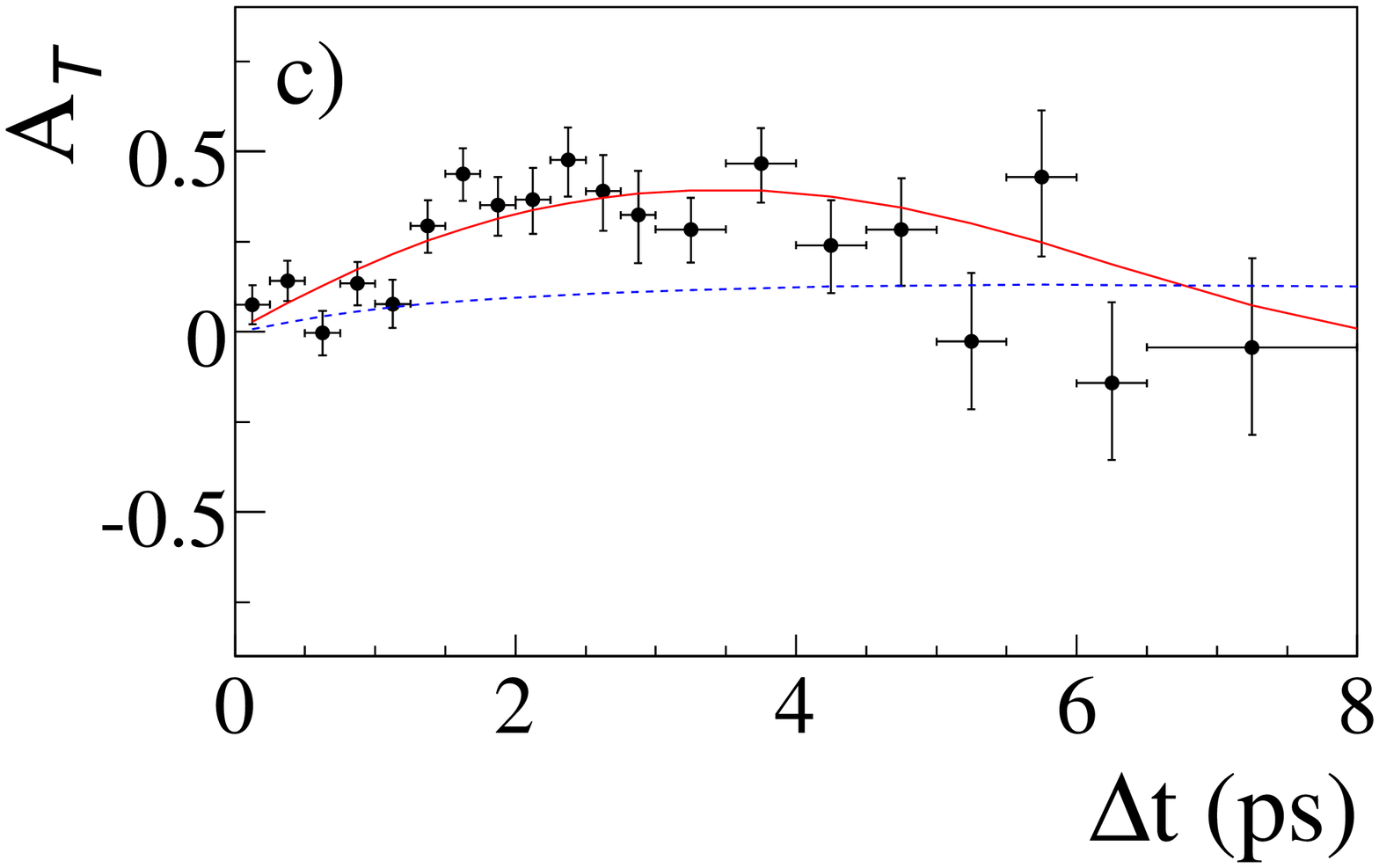} & 
  \includegraphics[width=0.23\textwidth]{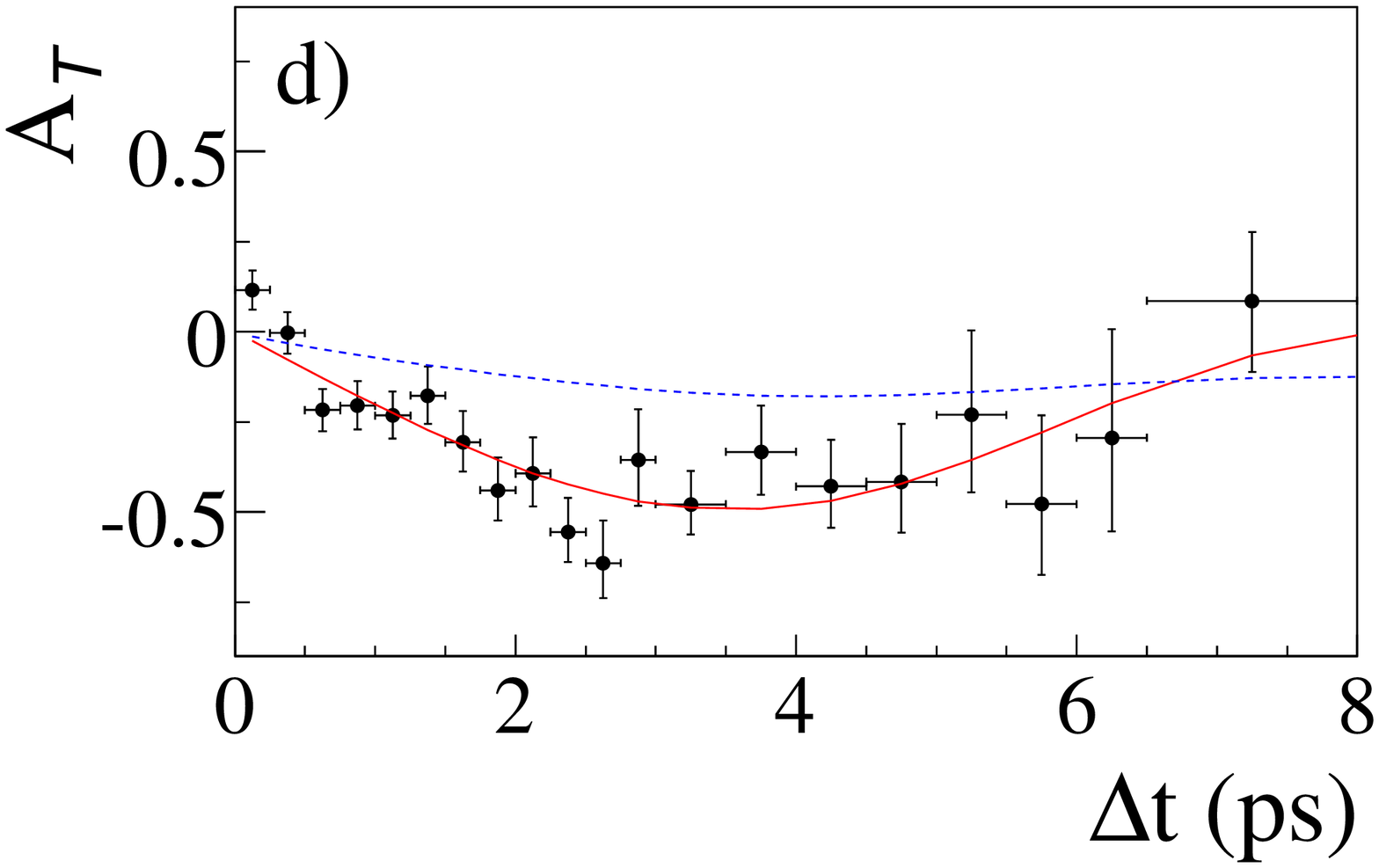} \\ 
\end{tabular} 
\caption{(color online). 
The four independent \T-violating asymmetries 
for 
transition 
a) $\Bzb \to \Bminus$ $(\ellp X,\ccbar\KS)$, 
b) $\Bplus \to \Bz$ $(\ccbar\KS, \ellp X)$, 
c) $\Bzb \to \Bplus$ $(\ellp X,\jpsi\KL)$, 
d) $\Bminus \to \Bz$ $(\jpsi\KL,\ellp X)$, 
for combined flavor categories with low misID  
(leptons and kaons), 
in the signal region  
($5.27< \mes < 5.29$~\gevcc for $\ccbar\KS$ modes and $|\de|<10$~\mev for $\jpsi\KL$).  
The points with error bars represent the data, the red solid and dashed blue curves represent the projections of the best  
fit results with and without \T violation, respectively. 
} 
\label{fig:AT} 
\end{center} 
\end{figure}

We build time-dependent asymmetries $A_{\T}(\dt)$ to visually demonstrate the \T-violating effect. 
For 
transition $\Bzb \to \Bminus$,  
\begin{equation} 
\label{eq:AT} 
A_{\T}(\dt) \equiv \dfrac{ {\cal H}_{\ellm,\KL}^-(\dt) - {\cal H}_{\ellp,\KS}^+(\dt) } 
                   { {\cal H}_{\ellm,\KL}^-(\dt) + {\cal H}_{\ellp,\KS}^+(\dt) }~, 
\end{equation} 
where ${\cal H}_{\alpha,\beta}^\pm(\dt) = {\cal H}_{\alpha,\beta}(\pm \dt) H(\dt)$.  
With this construction, $A_{\T}(\dt)$ is defined only for positive \dt values. 
Neglecting  
reconstruction effects,  
$A_{\T}(\dt)\approx \frac{\DeltaSpT}{2} \sin(\dmd\dt) + \frac{\DeltaCpT}{2} \cos(\dmd\dt)$. 
We  
introduce 
the other three \T-violating asymmetries similarly. 
Figure~\ref{fig:AT} shows the four  
observed 
asymmetries, 
overlaid with the projection of the best fit results  
to the \dt distributions with and without the eight \T-invariance restrictions: 
$\DeltaSpmT=\DeltaCpmT=0$, $\DeltaSpmCP=\DeltaSpmCPT$, and $\DeltaCpmCP=\DeltaCpmCPT$~\cite{ref:epaps}.

Using large samples of MC simulated data, we determine that the asymmetry parameters are unbiased and  
have Gaussian errors. 
Splitting the data by flavor category or data-taking period give consistent results. 
Fitting a single pair of $(S,C)$ coefficients, 
reversing the sign of $S$ under $\dt \leftrightarrow -\dt$, 
or $\Bplus \leftrightarrow \Bminus$ or $\Bz \leftrightarrow \Bzb$ exchanges, 
and the sign of $C$ under $\Bz \leftrightarrow \Bzb$ exchange, 
we obtain identical results to those obtained in  
Ref.~\cite{ref:Aubert:2009yr}. 
Performing the analysis with \B decays to $\ccbar\Kpm$ and $\jpsi\Kstarpm$ final states instead of the  
signal $\ccbar\KS$ and $\jpsi\KL$, respectively,  
we find that all the asymmetry parameters are consistent with zero. 
%

In evaluating systematic uncertainties in the asymmetry parameters, we follow the same procedure as in Ref.~\cite{ref:Aubert:2009yr}, 
with small changes~\cite{ref:epaps}. 
We considered the statistical uncertainties on the flavor misID probabilities, \dt resolution function, and \mes parameters. 
Differences in the misID  
probabilities 
and \dt resolution function between \Bflav and \CP final states, 
uncertainties due to assumptions in the resolution for signal and background components, 
compositions of the signal and backgrounds, the \mes and \de \pdfs, and the branching fractions for the  
backgrounds and their \CP properties, 
have also been accounted for. 
We also assign a systematic uncertainty corresponding to any  
deviation of the fit for MC simulated asymmetry parameters from their generated MC values, 
taking the largest between 
the deviation and its statistical uncertainty. 
Other sources of uncertainty such as our limited knowledge of $\Gamma_d$, \dmd, 
and other fixed parameters, 
the  
interaction region, 
the detector alignment,  
and effects due to a nonzero $\Delta\Gamma_d$ value in the time dependence and the normalization of the \pdf, 
are also considered. 
Treating $\ccbar \KS$ and $\jpsi \KL$ as orthogonal states and neglecting \CP violation for flavor categories without leptons, 
has an impact 
well below the statistical uncertainty. 
The total systematic uncertainties are shown in Table~\ref{tab:results}~\cite{ref:epaps}.

\begin{figure}[htb] 
\begin{center} 
\includegraphics[width=0.42\textwidth]{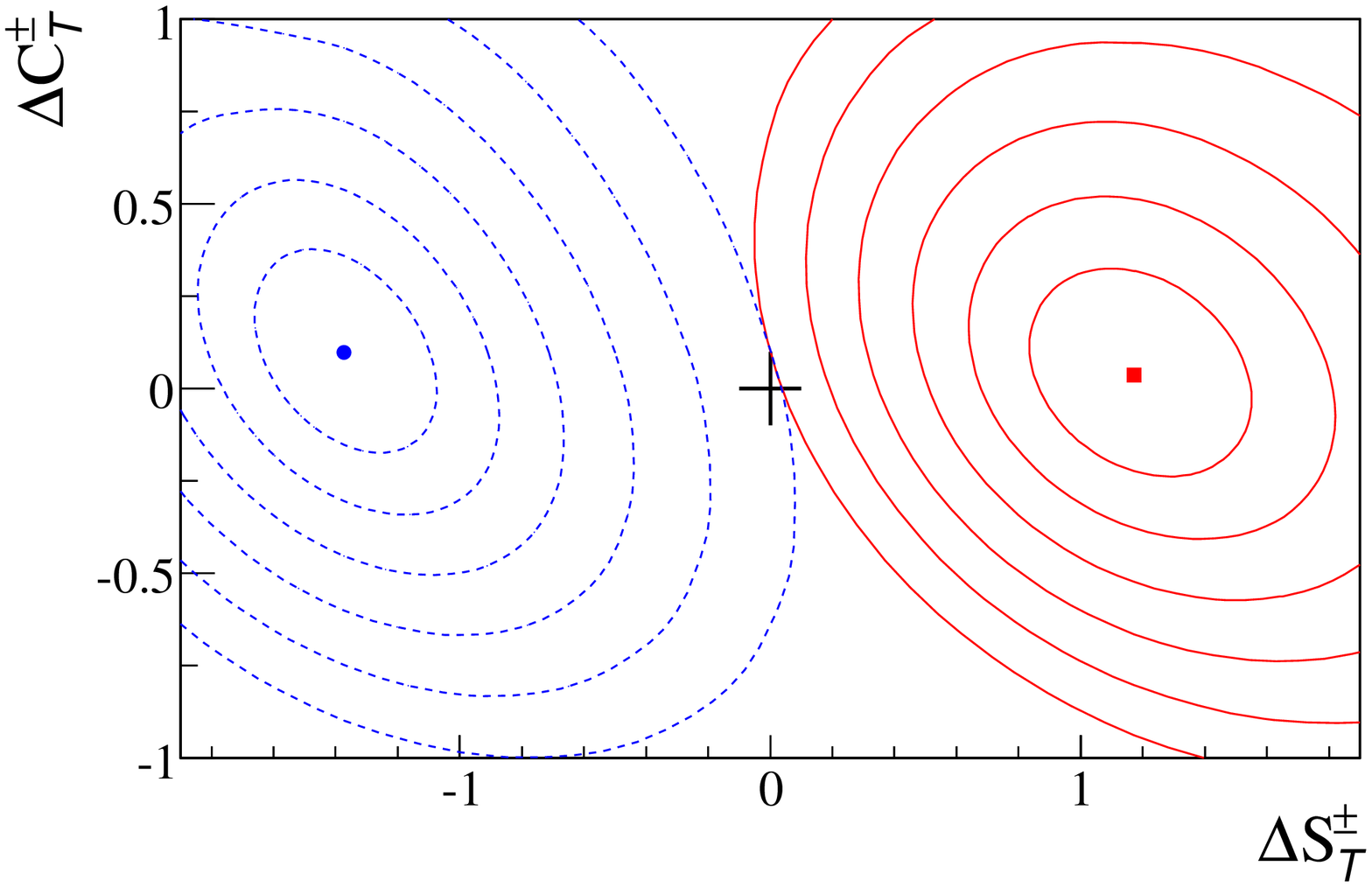} 
\caption{\label{fig:ContoursT} (color online). 
The central values (blue point and red square) and two-dimensional \CL contours 
for $1-\CL=0.317$, $4.55\times10^{-2}$, $2.70\times10^{-3}$, $6.33\times10^{-5}$,  
$5.73\times10^{-7}$, and $1.97\times10^{-9}$, 
calculated from the change in the value of \twoDLL 
compared with its value at maximum 
($\twoDLL=2.3,6.2,11.8,19.3,28.7,40.1$), 
for the pairs of \T-asymmetry parameters  
$(\DeltaSpT,\DeltaCpT)$ (blue dashed curves) and $(\DeltaSmT,\DeltaCmT)$ (red solid curves). 
Systematic uncertainties are included.  
The \T-invariance point is shown as a $+$ sign. 
} 
\end{center} 
\end{figure}

The significance of the \T-violation signal is evaluated based on the  
change in  
log-likelihood with respect to the  
maximum 
(\twoDLL). 
We  
reduce \twoDLL by a factor  
$1+\max \{m_{i}^2\}=1.61$ 
to account for systematic errors 
in the evaluation of the significance. 
Here,  
$m_{i}^2 = - 2( \ln {\cal L}_{i} - \ln {\cal L} )/s^2$,  
where $\ln {\cal L}$ is the maximum log-likelihood, 
$\ln {\cal L}_{i}$ 
is the log-likelihood with asymmetry parameter $i$ fixed to its  
total systematic variation  
and maximized over all other parameters, 
and $s^2 \approx 1$ is the  
change in $2\ln {\cal L}$ at $68\%$ confidence level (\CL) for one degree of freedom 
(d.o.f). 
Figure~\ref{fig:ContoursT} shows \CL contours calculated from the change \twoDLL 
in two dimensions for the \T-asymmetry parameters $(\DeltaSpT,\DeltaCpT)$ and $(\DeltaSmT,\DeltaCmT)$. 
The difference in the value of $2\ln {\cal L}$ at the best fit solution 
with and without \T violation is  
$226$ 
with eight d.o.f., 
including systematic uncertainties.  
Assuming Gaussian errors, this corresponds to a significance equivalent to $14$ standard deviations ($\sigma$), 
and thus constitutes direct observation of \T violation. The significance of \CP and \CPT violation 
is determined analogously, obtaining  
$307$ 
and  
$5$, 
respectively, 
equivalent  
to $17\sigma$  
and $0.3\sigma$,  
consistent with \CP violation and \CPT invariance.

In summary, we have measured \T-violating parameters in the time evolution  
of neutral \B mesons,  
by comparing  
the probabilities of  
$\Bzb \to \Bminus$, $\Bplus \to \Bz$, $\Bzb \to \Bplus$, and $\Bminus \to \Bz$ transitions, 
to their \T conjugate. 
We determine for the main \T-violating parameters 
$\DeltaSpT=-1.37 \pm 0.14~({\rm stat.}) \pm 0.06~({\rm syst.})$  
and  
$\DeltaSmT=1.17 \pm 0.18~({\rm stat.}) \pm 0.11~({\rm syst.})$, 
and observe directly for the first time 
a departure from \T invariance in the \B meson system,  
with a significance equivalent to $14\sigma$. 
Our results are consistent with current \CP-violating measurements  
obtained invoking \CPT invariance. 
They constitute the first 
observation of \T violation in any system 
through the exchange of initial and final states in transitions that can only be  
connected by a \T-symmetry transformation.

%% file: acknow_PRL.tex
We are grateful for the excellent luminosity and machine conditions 
provided by our \pep2\ colleagues,  
and for the substantial dedicated effort from 
the computing organizations that support \babar. 
The collaborating institutions wish to thank  
SLAC for its support and kind hospitality.  
This work is supported by 
DOE 
and NSF (USA), 
NSERC (Canada), 
CEA and 
CNRS-IN2P3 
(France), 
BMBF and DFG 
(Germany), 
INFN (Italy), 
FOM (The Netherlands), 
NFR (Norway), 
MES (Russia), 
MINECO (Spain), 
STFC (United Kingdom).  
Individuals have received support from the 
Marie Curie EIF (European Union), 
the A.~P.~Sloan Foundation (USA) 
and the Binational Science Foundation (USA-Israel).

%% file: document-extra.tex
\onecolumngrid 
\newpage 
 
\setcounter{page}{1} 
\setcounter{table}{0} 
\setcounter{figure}{0} 
 
\begin{center} 
{\large \bfseries \boldmath 
Observation of Time Reversal Violation in the \Bz Meson System}\\ 
The \babar\ Collaboration 
\end{center}

\begin{center} 
The following includes supplementary material for the Electronic 
Physics Auxiliary Publication Service.  
\end{center}

\begin{table}[htb!] 
\begin{center} 
\caption{Breakdown of main systematic uncertainties on the \T-, \CP-, and \CPT-asymmetry parameters and the $(S_{\ellp,\KS}^\pm, C_{\ellp,\KS}^\pm)$ coefficients 
for $\Bzb\to\Bminus$ and $\Bplus\to\Bz$ transitions. The indices \ellp and \KS stand for reconstructed final states that  
identify the \B meson as \Bz and \Bminus, respectively.  
The first nine rows in each panel are evaluated using similar procedures as in Ref.~\cite{ref:Aubert:2009yr}. 
The tenth and eleventh rows ($\Delta\Gamma_d/\Gamma_d$ and \pdf normalization) are estimated by varying $\Delta\Gamma_d/\Gamma_d$ by $\pm2\%$, 
while the $\sinh(\Delta\Gamma\dtau)$ and $\cosh(\Delta\Gamma\dtau)$ coefficients of the most general time-dependent decay  
rate $g_{\alpha,\beta}^\pm(\dtau)$~\cite{ref:method2012} are changed around their reference model values, 0 and 1, respectively. The \pdf normalization 
also accounts for systematic effects related to the \dt range used to normalize the \pdf. 
The total systematic uncertainty (last row in each panel) is calculated adding the individual systematic uncertainties in quadrature.  
\label{tab:systematics1to8}} 
\begin{ruledtabular} 
\begin{tabular}{l c c c c c c c c } 
\trule Systematic source & \DeltaSpT & \DeltaSmT & \DeltaCpT & \DeltaCmT & \DeltaSpCP & \DeltaSmCP & \DeltaCpCP & \DeltaCmCP  \\  [0.04in] 
\hline 
\trule Interaction region & 0.011 & 0.035 & 0.02 & 0.029 & 0.012 & 0.024 & 0.015 & 0.026 \\ 
\trule Flavor misID probabilities & 0.022 & 0.042 & 0.022 & 0.022 & 0.016 & 0.040 & 0.020 & 0.020 \\ 
\trule \dt resolution & 0.030 & 0.050 & 0.048 & 0.062 & 0.057 & 0.033 & 0.012 & 0.011 \\ 
\trule $\jpsi\KL$ background & 0.033 & 0.038 & 0.052 & 0.010 & 0.002 & 0.001 & 0.001 & 0.002 \\ 
\trule Background fractions and \CP content & 0.029 & 0.021 & 0.020 & 0.026 & 0.013 & 0.012 & 0.008 & 0.009 \\ 
\trule \mes parameterization & 0.011 & 0.002 & 0.005 & 0.002 & 0.016 & 0.008 & 0.005 & 0.004 \\ 
\trule $\Gamma_d$ and \dmd & 0.001 & 0.005 & 0.011 & 0.008 & 0.003 & 0.007 & 0.011 & 0.012 \\ 
\trule \CP violation for flavor ID categories & 0.018 & 0.019 & 0.001 & 0.001 & 0.009 & 0.008 & 0.006 & 0.006 \\ 
\trule Fit bias & 0.010 & 0.072 & 0.013 & 0.010 & 0.010 & 0.007 & 0.007 & 0.014 \\ 
\trule $\Delta\Gamma_d/\Gamma_d$ & 0.004 & 0.003 & 0.002 & 0.002 & 0.004 & 0.003 & 0.001 & 0.001 \\ 
\trule \pdf normalization & 0.013 & 0.019 & 0.005 & 0.004 & 0.017 & 0.012 & 0.006 & 0.007 \\ 
\hline 
\trule  Total & 0.064 & 0.112 & 0.08 & 0.077 & 0.068 & 0.061 & 0.033 & 0.041 \\ 
\hline \hline 
\\ 
\\ 
\hline \hline 
\trule Systematic source & \DeltaSpCPT & \DeltaSmCPT & \DeltaCpCPT & \DeltaCmCPT & \SpBzKs & \SmBzKs & \CpBzKs & \CmBzKs  \\ [0.05in] 
\hline 
\trule Interaction region & 0.015 & 0.024 & 0.023 & 0.026 & 0.014 & 0.009 & 0.015 & 0.008 \\ 
\trule Flavor misID probabilities & 0.018 & 0.008 & 0.009 & 0.009 & 0.013 & 0.020 & 0.012 & 0.010 \\ 
\trule \dt resolution & 0.062 & 0.033 & 0.051 & 0.072 & 0.051 & 0.030 & 0.045 & 0.012 \\ 
\trule $\jpsi\KL$ background & 0.046 & 0.021 & 0.029 & 0.015 & 0.002 & 0.001 & 0.001 & 0.001 \\ 
\trule Background fractions and \CP content & 0.024 & 0.020 & 0.024 & 0.016 & 0.012 & 0.004 & 0.007 & 0.007 \\ 
\trule \mes parameterization & 0.011 & 0.002 & 0.005 & 0.002 & 0.011 & 0.002 & 0.005 & 0.002 \\ 
\trule $\Gamma_d$ and \dmd & 0.004 & 0.001 & 0.002 & 0.003 & 0.003 & 0.003 & 0.009 & 0.008 \\ 
\trule \CP violation for flavor ID categories & 0.026 & 0.010 & 0.007 & 0.005 & 0.014 & 0.005 & 0.003 & 0.002 \\ 
\trule Fit bias & 0.018 & 0.026 & 0.007 & 0.021 & 0.005 & 0.017 & 0.006 & 0.015 \\ 
\trule $\Delta\Gamma_d/\Gamma_d$ & 0.003 & 0.002 & 0.002 & 0.001 & 0.002 & 0.001 & 0.001 & 0.001 \\ 
\trule \pdf normalization & 0.019 & 0.015 & 0.007 & 0.004 & 0.008 & 0.002 & 0.003 & 0.003 \\ 
\hline 
\trule  Total & 0.092 & 0.058 & 0.067 & 0.083 & 0.059 & 0.041 & 0.051 & 0.026 \\ 
\end{tabular} 
\end{ruledtabular} 
\end{center} 
\end{table}

\begin{figure}[htb] 
\begin{center} 
\begin{tabular}{cc} 
  \includegraphics[width=0.35\textwidth]{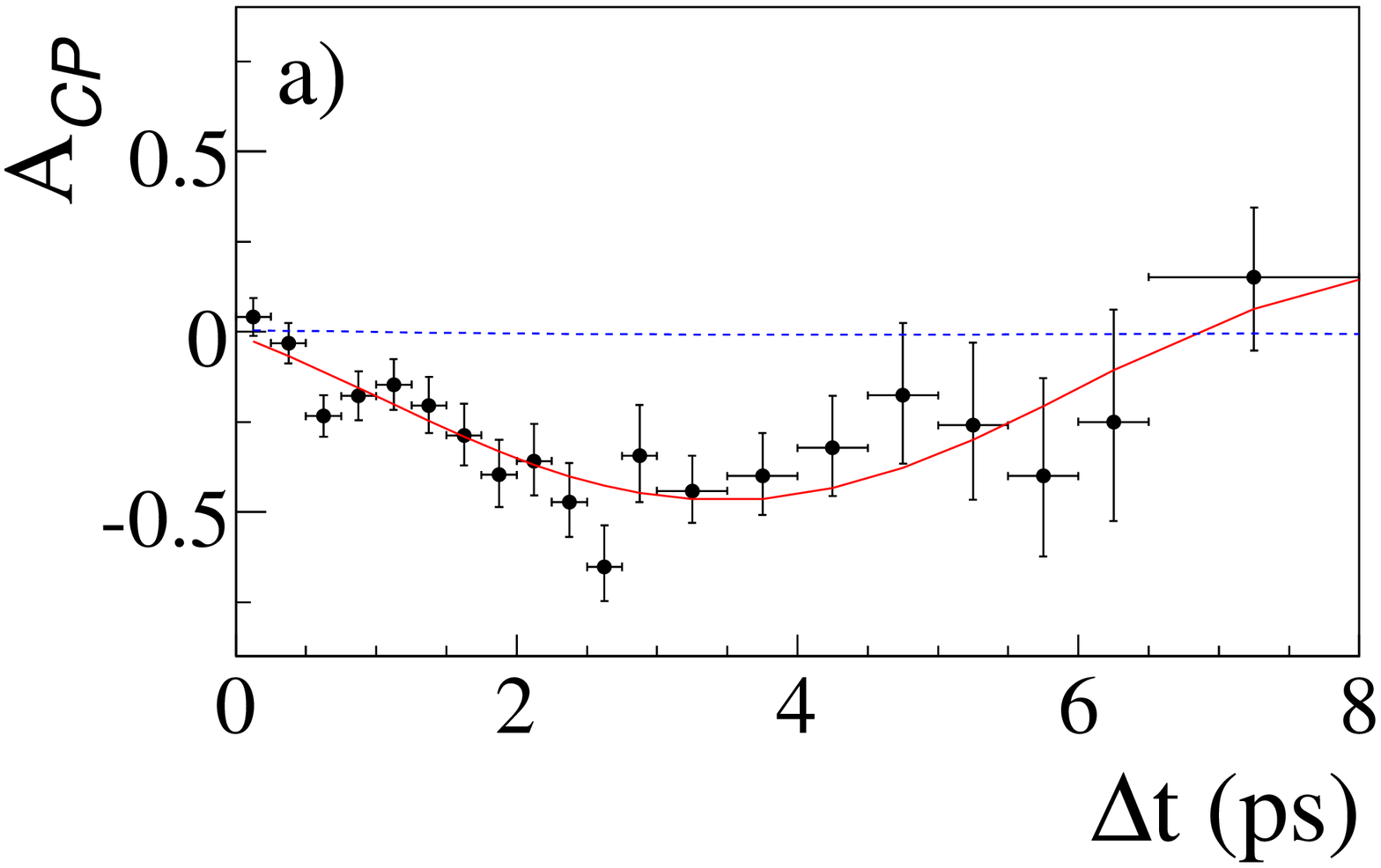} & 
  \includegraphics[width=0.35\textwidth]{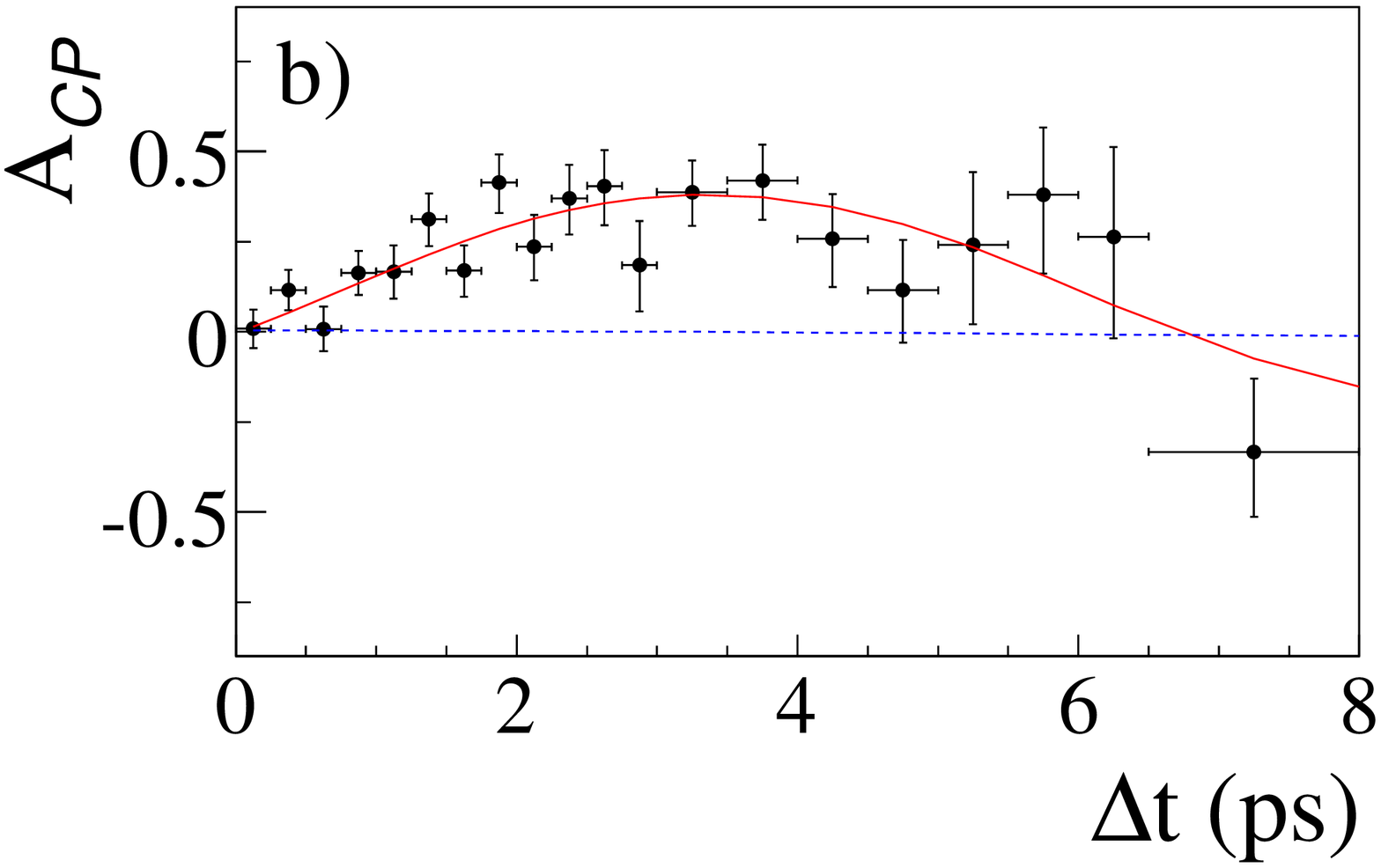} \\  
  \includegraphics[width=0.35\textwidth]{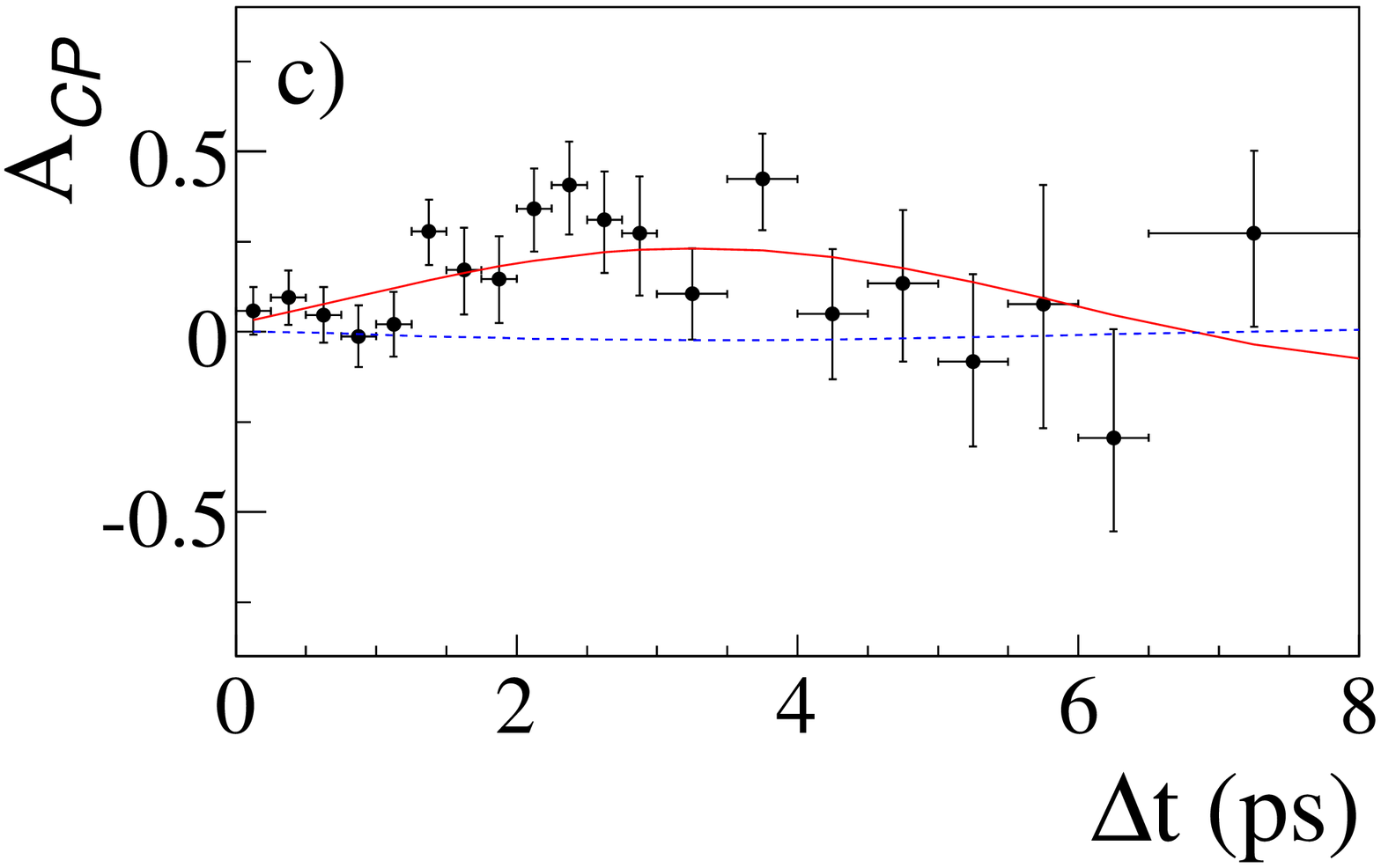} & 
  \includegraphics[width=0.35\textwidth]{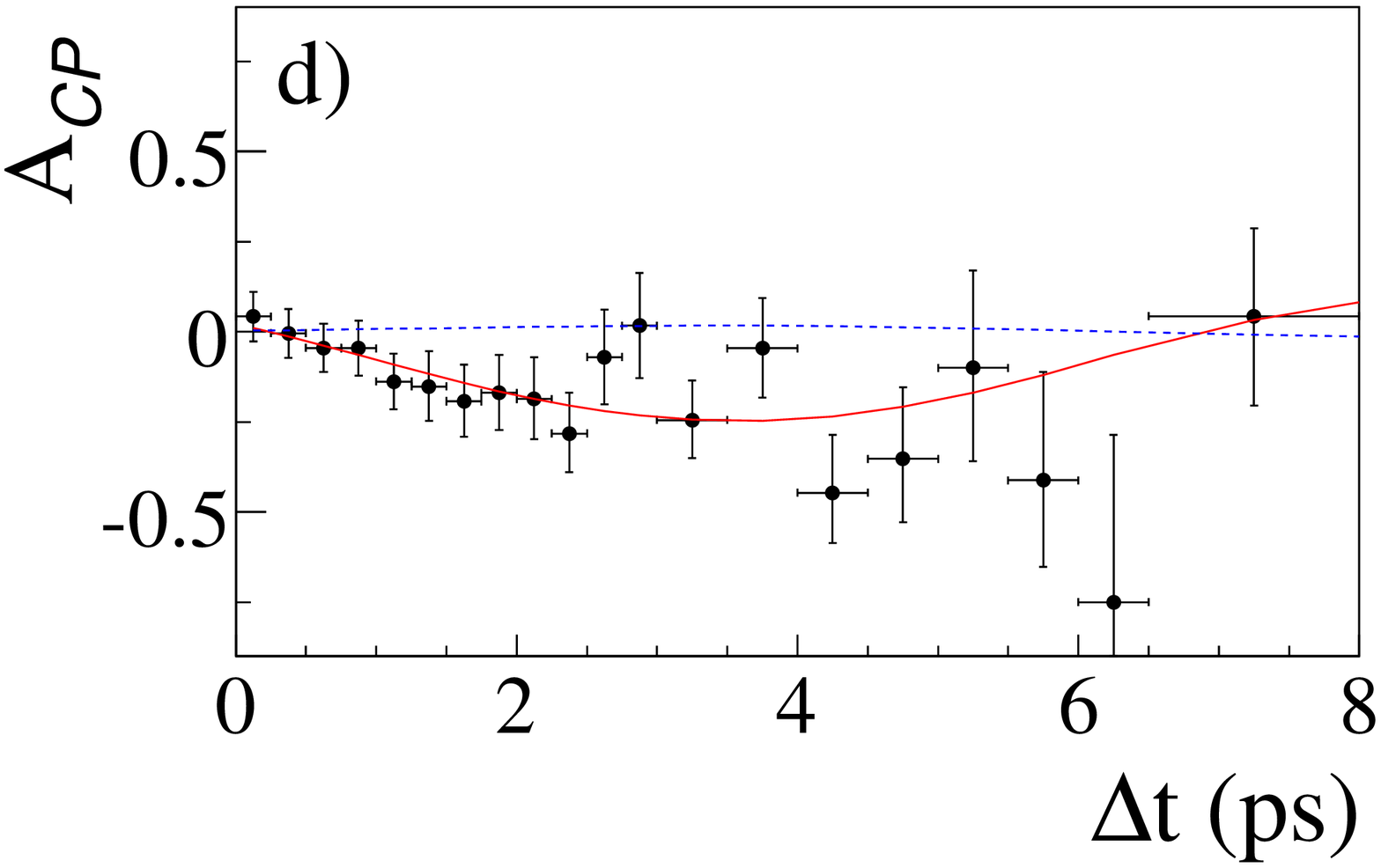} \\ 
\end{tabular} 
\caption{(color online). 
The four independent \CP-violating asymmetries 
for transition 
a) $\Bzb \to \Bminus$ $(\ellp X,\ccbar\KS)$, 
b) $\Bplus \to \Bz$ $(\ccbar\KS, \ellp X)$, 
c) $\Bzb \to \Bplus$ $(\ellp X,\jpsi\KL)$, 
d) $\Bminus \to \Bz$ $(\jpsi\KL,\ellp X)$, 
for combined flavor categories with low misID (leptons and kaons), in the signal region  
($5.27< \mes < 5.29$~\gevcc for $\ccbar\KS$ modes and $|\de|<10$~\mev for $\jpsi\KL$).  
The points with error bars represent the data, the red solid and dashed blue curves represent the projections of the best  
fit results with and without \CP violation, respectively. 
} 
\label{fig:ACP} 
\end{center} 
\end{figure}

\begin{figure}[htb] 
\begin{center} 
\begin{tabular}{cc} 
  \includegraphics[width=0.35\textwidth]{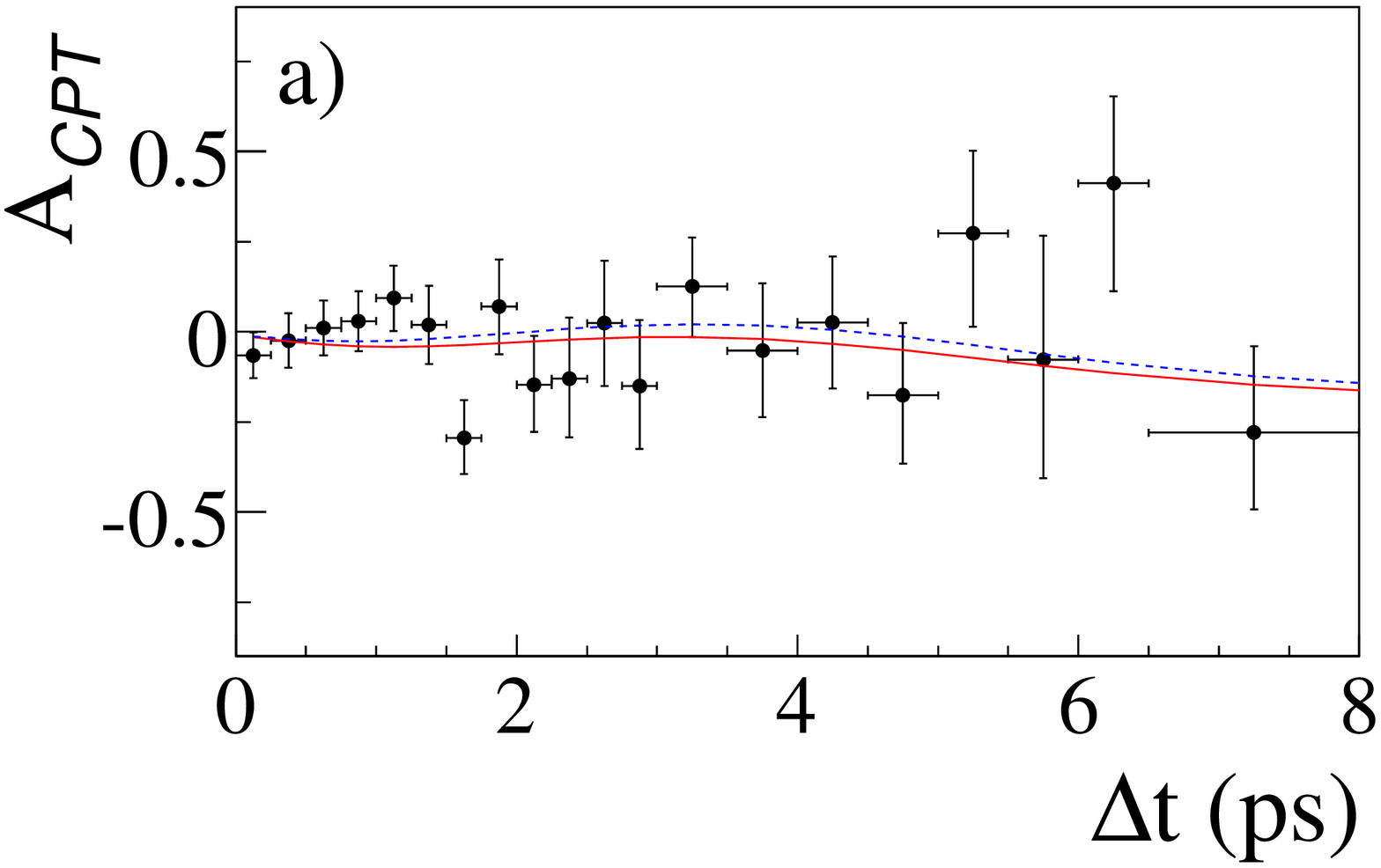} & 
  \includegraphics[width=0.35\textwidth]{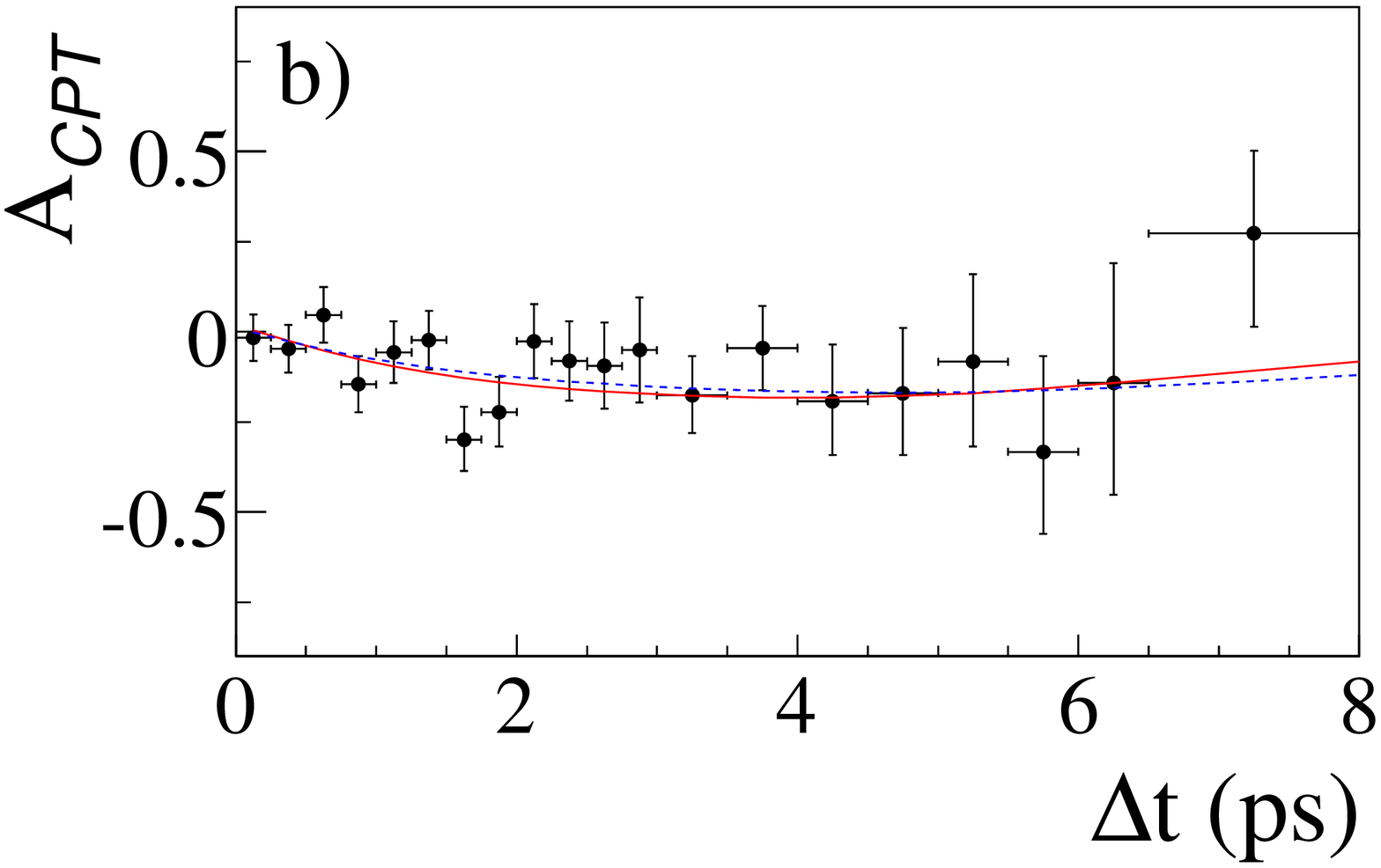} \\ 
  \includegraphics[width=0.35\textwidth]{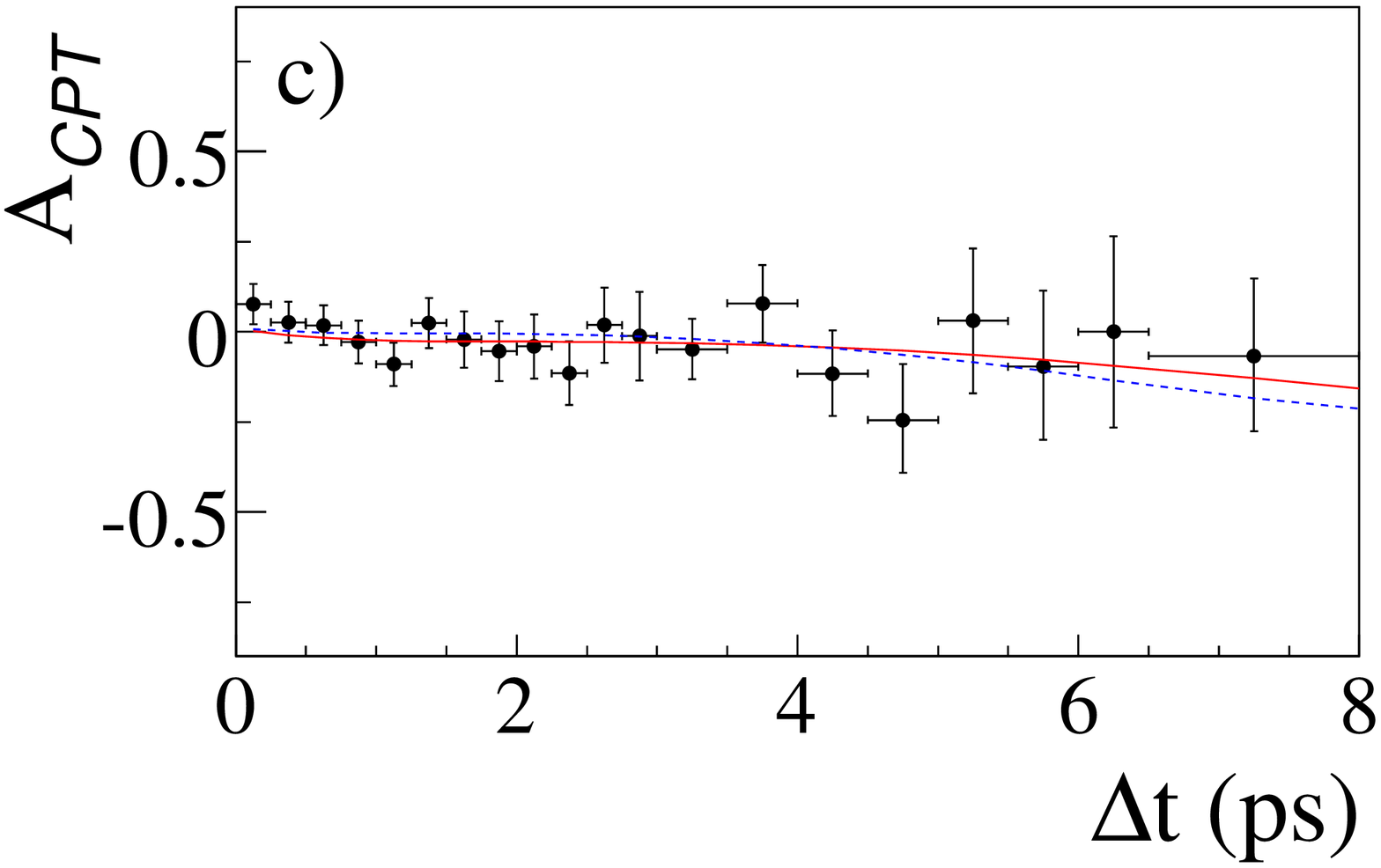} & 
  \includegraphics[width=0.35\textwidth]{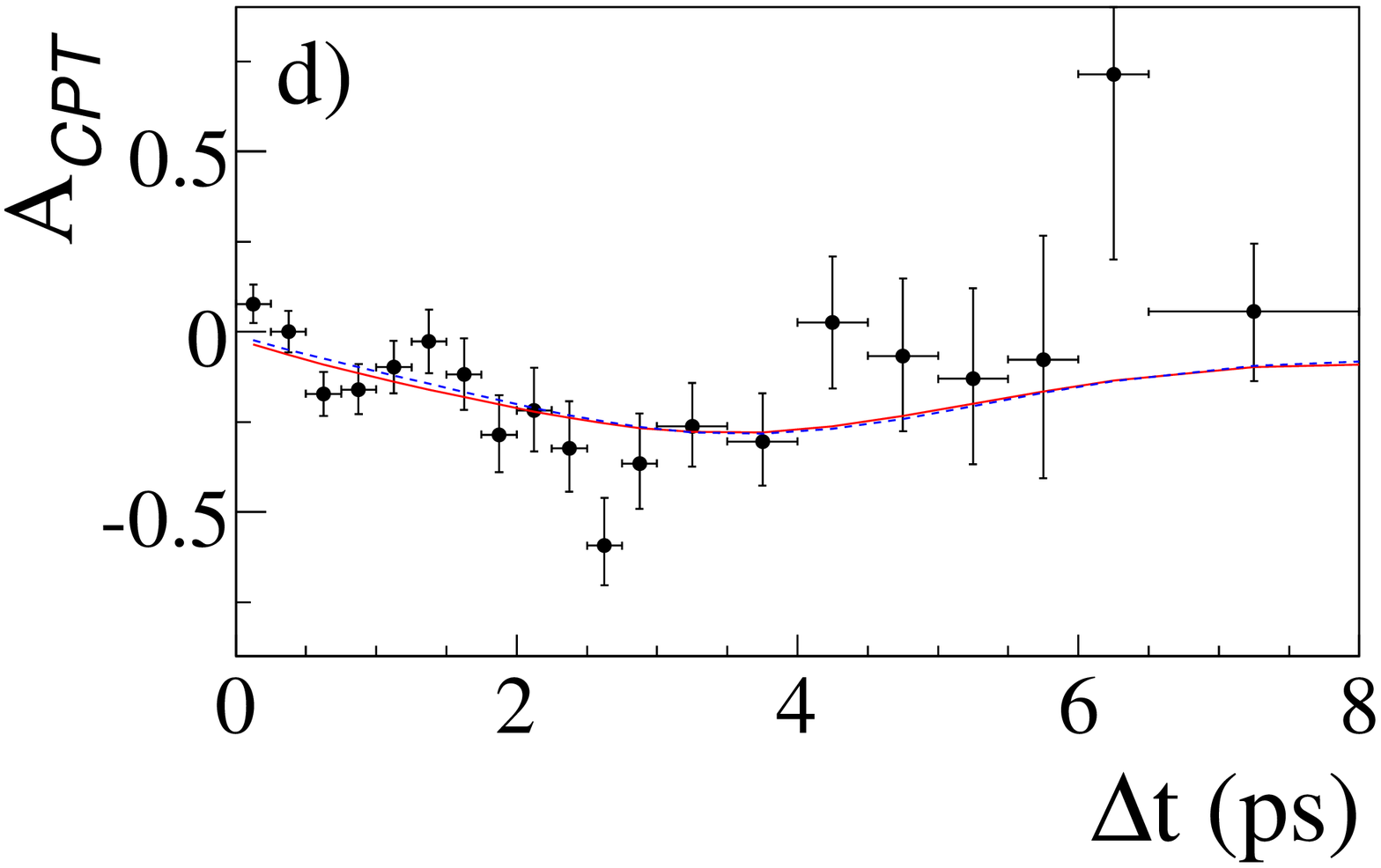} \\ 
\end{tabular} 
\caption{(color online). 
The four independent \CPT-violating asymmetries 
for transition 
a) $\Bplus \to \Bz$ $(\ccbar\KS,\ellp X)$, 
b) $\Bplus \to \Bzb$ $(\ccbar\KS,\ellm X)$, 
c) $\Bminus \to \Bz$ $(\jpsi\KL,\ellp X)$, 
d) $\Bminus \to \Bzb$ $(\jpsi\KL,\ellm X)$, 
for combined flavor categories with low misID (leptons and kaons), in the signal region  
($5.27< \mes < 5.29$~\gevcc for $\ccbar\KS$ modes and $|\de|<10$~\mev for $\jpsi\KL$).  
The points with error bars represent the data, the red solid and dashed blue curves represent the projections of the best  
fit results with and without \CPT violation, respectively. 
} 
\label{fig:ACPT} 
\end{center} 
\end{figure}

\begin{figure}[htb] 
\begin{center} 
\begin{tabular}{cc} 
\includegraphics[width=0.42\textwidth]{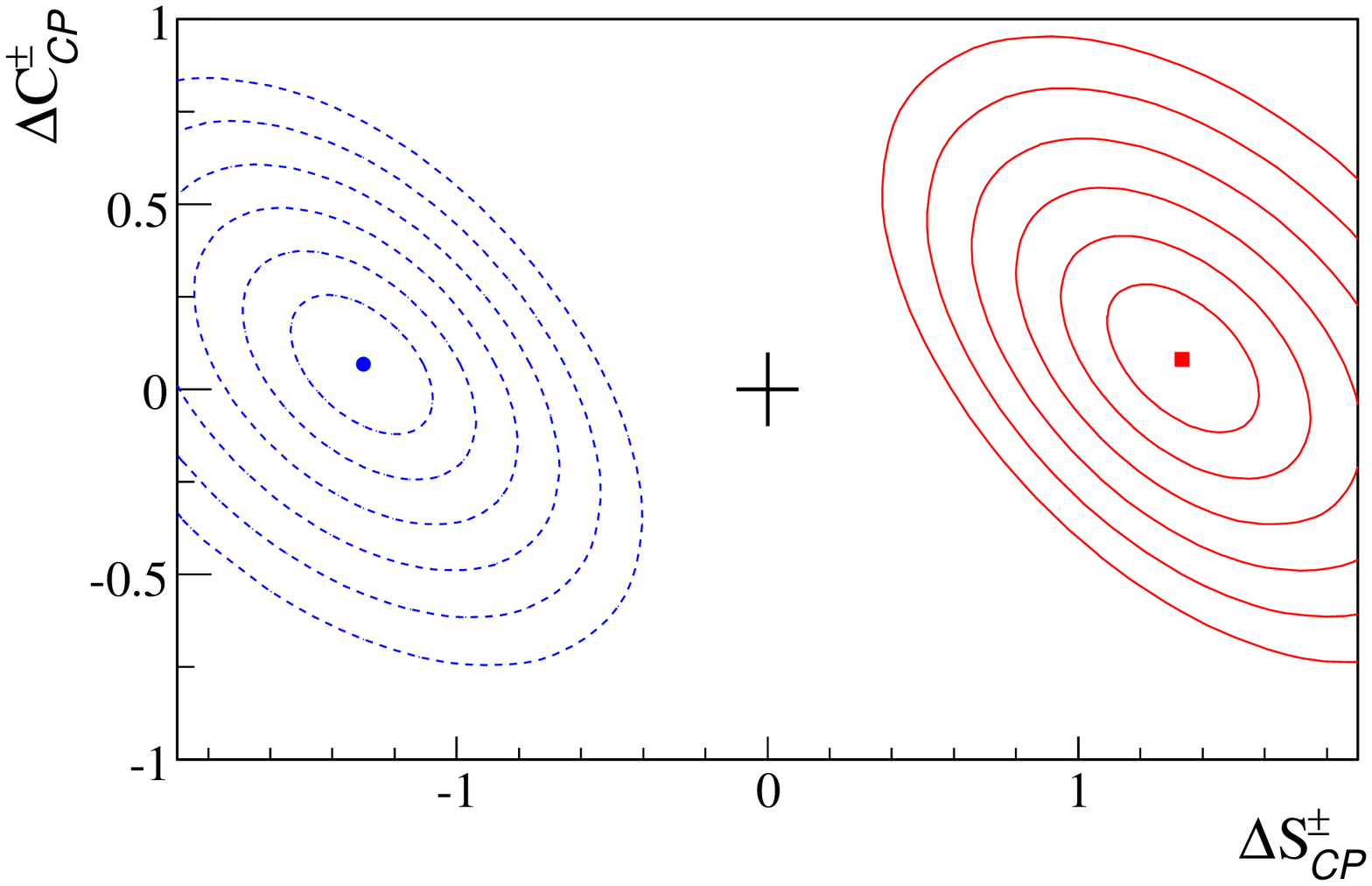} & 
\includegraphics[width=0.42\textwidth]{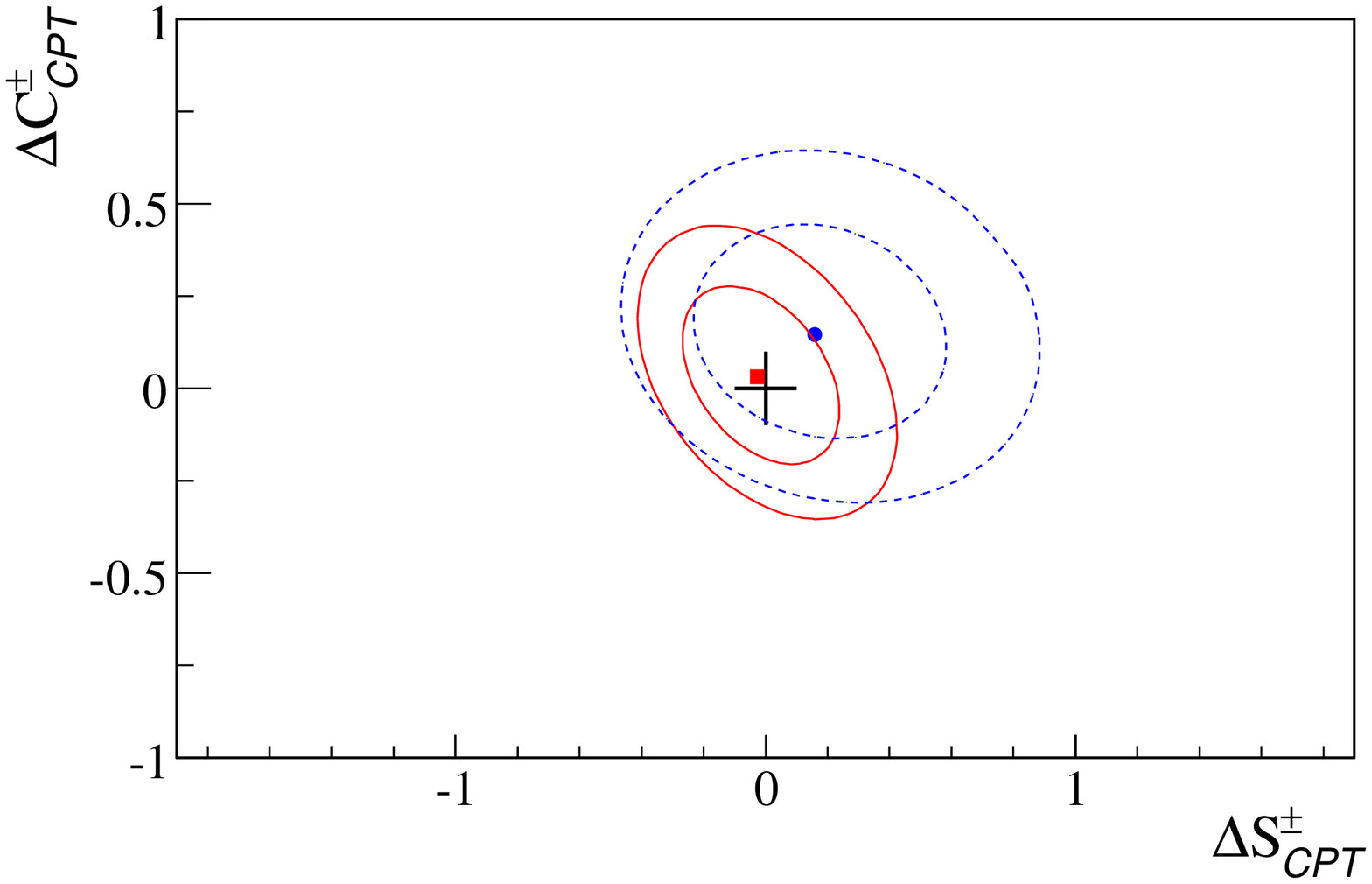} 
\end{tabular} 
\caption{\label{fig:ContoursT} (color online). 
The central values (blue point and red square) and two-dimensional \CL contours 
for $1-\CL=0.317$, $4.55\times10^{-2}$, $2.70\times10^{-3}$, $6.33\times10^{-5}$,  
$5.73\times10^{-7}$, and $1.97\times10^{-9}$,  
calculated from the change in the value of \twoDLL 
compared with its value at maximum, for the pairs of \CP- (left) and \CPT- (right) asymmetry parameters  
$(\DeltaSpCP,\DeltaCpCP)$ and $(\DeltaSpCPT,\DeltaCpCPT)$ (blue dashed curves)  
and 
$(\DeltaSmCP,\DeltaCmCP)$, $(\DeltaSmCPT,\DeltaCmCPT)$ (red solid curves).  
Systematic uncertainties are included. The \CP- and \CPT-invariance points are shown as a plus sign ($+$). 
} 
\end{center} 
\end{figure}

  \begin{table}[htb!] 
    \begin{center} 
      \caption{Measured values of the $(S_{\alpha,\beta}^\pm, C_{\alpha,\beta}^\pm)$ coefficients. 
The first uncertainty is statistical and the second systematic. 
The indices $\alpha=\ellm, \ellp$ and $\beta=\KS,\KL$ stand for reconstructed final states that  
identify the \B meson as \Bzb, \Bz and \Bminus, \Bplus, respectively. 
\label{tab:SCresults}}        
\begin{ruledtabular} 
\begin{tabular}{ r  l  c  c }		 
\multicolumn{2}{c}{Transition}                 & ~~~Parameter~~~ & Result \\ \hline 
\trule $\Bminus\to\Bzb$  & $(\jpsi\KL,\ellm X)$  & \SmBzbKl & $-0.83\pm0.11\pm0.06$ \\ 
\trule $\Bz\to\Bminus $  & $(\ellm X,\ccbar\KS)$ & \SpBzbKs & $-0.76\pm0.06\pm0.04$ \\ 
\trule $\Bminus\to\Bz $  & $(\jpsi\KL,\ellp X)$  & \SmBzKl  & $\phm0.70\pm0.19\pm0.12$ \\ 
\trule $\Bzb\to\Bminus$  & $(\ellp X,\ccbar\KS)$ & \SpBzKs  & $\phm0.55\pm0.09\pm0.06$ \\ 
\trule $\Bz\to\Bplus  $  & $(\ellm X,\jpsi\KL)$  & \SpBzbKl & $\phm0.51\pm0.17\pm0.11$ \\ 
\trule $\Bplus\to\Bzb $  & $(\ccbar\KS,\ellm X)$ & \SmBzbKs & $\phm0.67\pm0.10\pm0.08$ \\ 
\trule $\Bzb\to\Bplus $  & $(\ellp X,\jpsi\KL)$  & \SpBzKl  & $-0.69\pm0.11\pm0.04$ \\ 
\trule $\Bplus\to\Bz  $  & $(\ccbar\KS,\ellp X)$ & \SmBzKs  & $-0.66\pm0.06\pm0.04$ \\ [0.07in] \hline 
\trule $\Bminus\to\Bzb$  & $(\jpsi\KL,\ellm X)$  & \CmBzbKl & $\phm0.11\pm0.12\pm0.08$ \\ 
\trule $\Bz\to\Bminus $  & $(\ellm X,\ccbar\KS)$ & \CpBzbKs & $\phm0.08\pm0.06\pm0.06$ \\ 
\trule $\Bminus\to\Bz $  & $(\jpsi\KL,\ellp X)$  & \CmBzKl  & $\phm0.16\pm0.13\pm0.06$ \\ 
\trule $\Bzb\to\Bminus$  & $(\ellp X,\ccbar\KS)$ & \CpBzKs  & $\phm0.01\pm0.07\pm0.05$ \\ 
\trule $\Bz\to\Bplus  $  & $(\ellm X,\jpsi\KL)$  & \CpBzbKl & $-0.01\pm0.13\pm0.08$ \\ 
\trule $\Bplus\to\Bzb $  & $(\ccbar\KS,\ellm X)$ & \CmBzbKs & $\phm0.03\pm0.07\pm0.04$ \\ 
\trule $\Bzb\to\Bplus $  & $(\ellp X,\jpsi\KL)$  & \CpBzKl  & $-0.02\pm0.11\pm0.08$ \\ 
\trule $\Bplus\to\Bz  $  & $(\ccbar\KS,\ellp X)$ & \CmBzKs  & $-0.05\pm0.06\pm0.03$ \\ [0.07in] 
      \end{tabular} 
\end{ruledtabular}        
    \end{center}	 
  \end{table}

\begin{table}[!htb] 
\begin{center} 
\caption{\label{tab:rhostat}  
Statistical correlation coefficients for the vector of $(S_{\alpha,\beta}^\pm,C_{\alpha,\beta}^\pm)$ measurements  
given in the same order as 
in Table~\ref{tab:SCresults}. Only lower off-diagonal terms are written, in \%. 
} 
$\left( 
\begin{tabular}{rrrrrrrrrrrrrrrr}

\input SCcorrelation_stat

 
\end{tabular} 
\right)$ 
\end{center} 
 
\end{table}

\begin{table}[!htb] 
\caption{\label{tab:rhostat}  
Systematic correlation coefficients for the vector of $(S_{\alpha,\beta}^\pm,C_{\alpha,\beta}^\pm)$ measurements  
given in the same order as 
in Table~\ref{tab:SCresults}. Only lower off-diagonal terms are written, in \%. 
} 
\begin{center} 
$\left( 
\begin{tabular}{rrrrrrrrrrrrrrrr}

\input SCcorrelation_syst

\end{tabular} 
\right)$ 
\end{center} 
 
\end{table}

%% file: SCcorrelation_stat.tex
100 \\ 
0 & 100 \\ 
-14 & 0 & 100 \\ 
2 & -6 & 0 & 100 \\ 
8 & 0 & 41 & 0 & 100 \\ 
0 & 18 & 0 & 38 & 0 & 100 \\ 
6 & 0 & 19 & 0 & -7 & 0 & 100 \\ 
0 & 10 & 0 & 16 & 0 & -9 & 1 & 100 \\ 
-45 & 0 & 38 & -1 & 31 & 0 & 9 & 0 & 100 \\ 
0 & -33 & 0 & 31 & 0 & 28 & 0 & 6 & 0 & 100 \\ 
27 & 0 & -9 & 0 & 23 & 0 & 18 & 0 & -14 & 0 & 100 \\ 
0 & 28 & 0 & -14 & 0 & 23 & 0 & 18 & 1 & -15 & 0 & 100 \\ 
15 & 0 & 21 & 0 & -21 & 0 & 27 & 0 & -16 & 0 & 22 & 0 & 100 \\ 
0 & 18 & 0 & 21 & 0 & -18 & 0 & 29 & 0 & -16 & 0 & 21 & 0 & 100 \\ 
1 & 0 & 25 & 0 & 31 & 0 & -37 & 0 & 22 & 0 & -15 & 0 & -20 & 0 & 100 \\ 
0 & 7 & 0 & 23 & 0 & 31 & 0 & -41 & 0 & 20 & 0 & -17 & 0 & -20 & 0 & 100 \\

%% file: SCcorrelation_syst.tex
100 \\ 
6 & 100 \\ 
18 & -14 & 100 \\ 
44 & 3 & 66 & 100 \\ 
16 & -4 & 57 & 58 & 100 \\ 
37 & -19 & 67 & 66 & 44 & 100 \\ 
-5 & -5 & 10 & 8 & -4 & -3 & 100 \\ 
30 & -19 & 57 & 59 & 10 & 58 & 6 & 100 \\ 
-28 & -10 & 39 & 13 & 43 & 21 & -8 & -1 & 100 \\ 
42 & -20 & 60 & 68 & 57 & 72 & -6 & 47 & 30 & 100 \\ 
-31 & 0 & 23 & 17 & 20 & 8 & 11 & 6 & 58 & 18 & 100 \\ 
41 & -27 & 70 & 66 & 46 & 64 & 0 & 71 & 32 & 81 & 20 & 100 \\ 
31 & -16 & 63 & 63 & 39 & 67 & -23 & 59 & 39 & 63 & 24 & 73 & 100 \\ 
1 & -1 & 15 & 7 & 2 & 2 & -31 & 5 & 23 & 7 & 5 & 18 & 49 & 100 \\ 
28 & -23 & 73 & 72 & 52 & 61 & -1 & 64 & 43 & 69 & 28 & 84 & 83 & 39 & 100 \\ 
-14 & -13 & 12 & -6 & -34 & 11 & 2 & 34 & 23 & 0 & 31 & 17 & 26 & 15 & 15 & 100 \\

%% file: paper.bbl
\begin{thebibliography}{99} 
 
\bibitem{ref:christenson:1964} 
 J.H.~Christenson {\it et al.}, \jprl{13}, 138 (1964). 
 
\bibitem{ref:mixingInducedCP-Bs} B.~Aubert {\it et al.} (\babar\ Collaboration), \jprl{87}, 091801 (2001);  
                                 K.~Abe {\it et al.} (Belle Collaboration), \jprl{87}, 091802 (2001). 
\bibitem{ref:directCP-Bs} B.~Aubert {\it et al.} (\babar\ Collaboration), \jprl{93}, 131801 (2004);  
                          Y.~Chao {\it et al.} (Belle Collaboration), \jprl{93}, 191802 (2004). 
 
\bibitem{ref:CKM:1963:1973} 
 N.~Cabibbo, \jprl{10}, 531 (1963); M.~Kobayashi and T.~Maskawa, Prog. Theor. Phys. {\bf 49}, 652 (1973). 
 
 
\bibitem{ref:CPTtheorem}  
G.~L\"uders, Math. Fysik. Medd. Kgl. Danske Akad. Ved. Volume 28, 1954, p. 5; 
J.S.~Bell, Birmingham University thesis (1954); 
W.~Pauli, in W.~Pauli, ed., Niels~Bohr and the Development of Physics (McGraw-Hill,  
NY,  
1955). 
 
\bibitem{ref:CPTtests}  
R.~Carosi {\it et al.}, \plb{237}, 303 (1990);  
A.~Alavi-Harati {\it et al.}, \jprd{67}, 012005 (2003); 
B.~Schwingenheuer {\it et al.}, \jprl{74}, 4376 (1995). 
 
\bibitem{ref:TestsConservationLaws} See ``Tests of conservation laws'' review in~\cite{ref:pdg2010}. 
 
\bibitem{ref:pdg2010} K.~Nakamura~{\it et al.} (Particle Data Group), \jpg{37}, 075021 (2010). 
 
\bibitem{ref:Angelopoulos} A.~Angelopoulus {\it et al.} (CPLEAR Collaboration), \plb{444}, 43 (1998).         
 
\bibitem{ref:Kabir} P.~K.~Kabir, \jprd{2}, 540 (1970). 
 
\bibitem{ref:Wolfenstein} L.~Wolfenstein, \jprl{83}, 911 (1999). 
 
\bibitem{ref:Wolfenstein2} L.~Wolfenstein, \ijmp{8}, 501 (1999). 
 
\bibitem{ref:Gerber} H.J.~Gerber, \epjc{35}, 195 (2004), and references therein. 
 
\bibitem{ref:TviolationBs}  
B.~Aubert {\it et al.} (\babar\ Collaboration), \jprl{96}, 251802 (2006), \jprl{92}, 181801 (2004); 
E.~Nakano {\it et al.} (Belle Collaboration), \jprd{73}, 112002 (2006); 
V.M.~Abazov {\it et al.} (D0 Collaboration), \jprl{105}, 081801 (2010), \jprl{98}, 151801 (2007); 
F.~Abe {\it et al.} (CDF Collaboration), \jprd{55}, 2546 (2006). 
 
\bibitem{ref:edm} J.J.~Hudson {\it et al.}, Nature~{\bf 473}, 493-496 (2011); C.A.~Baker {\it et al.}, \jprl{97}, 131801 (2006). 
 
\bibitem{ref:method2012} 
  J.~Bernabeu, F.~Martinez-Vidal, P.~Villanueva-Perez, JHEP {\bf 1208}, 064 (2012). 
 
\bibitem{ref:bernabeuPLB-NPB} M.~C.~Ba\~nuls and J.~Bernabeu, \plb{464}, 117 (1999); \npb{590}, 19 (2000). 
 
\bibitem{ref:QuinnDiscrete} H.~R.~Quinn, J. Phys. Conf. Ser. {\bf 171}, 011001 (2009). 
 
\bibitem{ref:BernabeuDiscrete} J.~Bernabeu, J. Phys. Conf. Ser. {\bf 335}, 012011 (2011). 
 
\bibitem{ref:Aubert:2001tu} 
  B.~Aubert {\it et al.} (\babar\ Collaboration), 
  \nima{479}, 1 (2002). 
 
\bibitem{ref:Aubert:2009yr} 
  B.~Aubert {\it et al.} (\babar\ Collaboration), 
  \jprd{79}, 072009 (2009). 
 
 
\bibitem{ref:CPVreview} See 
``\CP violation in meson decays'' review in~\cite{ref:pdg2010}. 
 
 
 
\bibitem{ref:epaps} See supplementary material for  
                    breakdown of the main systematic uncertainties on the asymmetry parameters, 
                    \CP- and \CPT-violating asymmetries,   
                    and complete $(S_{\alpha,\beta}^\pm, C_{\alpha,\beta}^\pm)$ analysis results.  
 
 
\end{thebibliography}
